\documentclass[useAMS,usenatbib]{mn2e}
\usepackage{graphicx}
\begin{document}
\title[Metal Cooling and Diffusion]{The Enrichment of the
  Intergalactic Medium With Adiabatic Feedback I: Metal Cooling and Metal Diffusion}
\author[S.~Shen et al.]{S.~Shen,$^1$\thanks{Email:
    shens@physics.mcmaster.ca} J.~Wadsley $^1$ and G.~Stinson $^{1,2}$ \\
 $^1$ Department of Physics and Astronomy, McMaster University, Main
 Street West, Hamilton L8S 4M1, Canada\\
 $^2$ Jeremiah Horrocks Institute, University of Central Lancashire, Preston, PR1 2HE }
\maketitle 

\begin{abstract}

A study of metal enrichment of the intergalactic medium
(IGM) using a series of smooth particle hydrodynamics (SPH)
simulations is presented, employing models for metal cooling and the turbulent
diffusion of metals and thermal energy. An adiabatic feedback
  mechanism was adopted where gas cooling was prevented on the
  timescale of supernova bubble expansion to generate galactic winds
  without explicit wind particles.  The simulations produced a cosmic
  star formation history (SFH) that is broadly consistent with
  observations until z $\sim$ 0.5, and a steady evolution of the
  universal neutral hydrogen fraction ($\Omega_{\rm H I}$) that
  compares reasonably well with observations.   
The evolution of the mass and metallicities in stars and various gas phases
was investigated.  At z=0, about 40\% of the baryons are in the
warm-hot intergalactic medium  
(WHIM), but most metals (80\%-90\%) are locked in stars.  At higher
redshifts the proportion of metals in the IGM is higher due to more
efficient loss from galaxies.  The results also indicate that IGM metals
primarily reside in the WHIM throughout cosmic history,
which differs from simulations with hydrodynamically decoupled
explicit winds.      
The metallicity of the WHIM lies between 0.01 and 0.1 solar with a
slight decrease at lower redshifts.  The metallicity 
evolution of the gas inside galaxies are broadly consistent with
observations, but the diffuse IGM is under enriched at  z $\sim$ 2.5. 

Galactic winds most efficiently enrich the IGM for halos in the
intermediate mass range $10^{10}$M$_{\sun}$ - $10^{11}$ M$_{\sun}$. At
the low mass end gas is prevented from accreting onto halos and has very low
metallicities. At the high mass end,  the fraction of halo baryons escaped
as winds declines along with the decline of stellar mass
  fraction of the galaxies.  This is likely because of the 
decrease in star formation activity and decrease in wind escape
efficiency. Metals enhance cooling which allows WHIM gas to cool onto
galaxies and increases star formation.  Metal diffusion 
allows winds to mix prior to escape, decreasing the IGM metal content
in favour of gas within galactic halos and star forming 
gas.  Diffusion significantly increases the amount of gas with low
metallicities and changes the density-metallicity relation.

\end{abstract}

\section{Introduction} 

The intergalactic medium (IGM) contains most of the baryons in the
Universe and it provides the fuel for galaxies to form stars in which
metals are produced. In turn, supernovae and galactic winds enrich the
IGM with metals,  while stars and  active galactic nuclei (AGN) emit
UV photons.  This interplay between the IGM and galaxies, mediated by
metal cooling in the presence of UV, regulates the formation of stars in the
universe.  The evolution and enrichment history of the IGM provides a
record of this interplay. 
   
Observations of metal absorption lines (e.g., C III , C IV, Si
III, S IV and O VI) in quasar spectra show that the
intergalactic medium (IGM) far outside large galaxies
($\rho/\rho_{mean} < 10$) is enriched  
  \citep[e.g.][]{Songaila96, Dave98,
  Ellison00, Schaye00, Pettini03, Schaye03, Aguirre04,
  Simcoe04}. There is evidence for enrichment extending back to $z >
5$ \citep{Pettini03, Simcoe06}  
though metallicities do not evolve much from z=4  to z=2
\citep{Schaye03}.  Since metals are created in stars inside galaxies,
those in the IGM must have escaped from galaxies.   
Exactly how this occurs is unclear. It typically assumed that galactic
winds, driven by star formation, dominate IGM enrichment
\citep{Aguirre07}.   
Winds can be either launched as the ejecta of a large
number of co-existing supernova (SN) explosions \citep{Heckman90}, 
driven by the injection of momentum by SN and stellar winds or by radiation pressure
from starbursts and AGN \citep{Murray05}.
Observations of galactic winds \citep[e.g.,][]{Heckman01, Pettini01} have found
that galactic wind velocities range from hundreds to thousands of km/s 
and the mass loss rates are comparable to
star formation rates.  Wind material has a complex, multiphase
structure but most metals are expected to be entrained in the hot phase
\citep[][and references therein]{Veilleux05}.  
   
Although detailed hydrodynamical simulations of the interstellar
medium (ISM) in galaxies \citep[e.g,,][]{MacLow99, Strickland00,
Williams02} have been able to generate galactic outflows and have
explored various properties of winds and the gas dynamics in different
phases, current 
cosmological simulations lack the resolution to launch or track winds directly.
Hot, low density SN bubbles are unresolved in such simulations which initially led to an
 overcooling problem that produced
unrealistically concentrated simulated galaxies \citep{Navarro97}. As 
a result, various ``subgrid'' stellar feedback and wind models have emerged.  
These models serve two functions: to regulate star formation and the
properties of the ISM and to redistribute gas (and newly formed
metals) both within and into the environment around galaxies.  There
are three main approaches, energetic feedback, kinetic feedback and
modifications to the effective equation of state which behaves similarly
to an increased effective pressure. 

Energetic feedback in its simplest form involves simply adding the
stellar feedback as thermal energy, but this suffers from overcooling
\citep{Katz96}.  Kinetic feedback \citep[e.g.,][]{Navarro93, SH03, 
  Oppenheimer06, DallaVecchia08} converts part of the SN energy into
kinetic energy in the gas. The effectiveness of kinetic feedback is
strongly dependent on the resolution and hydrodynamic method. 

\citet{SH03} argued that regulated star formation creates an effective
pressure in the ISM and this was modelled directly in the \textsc{gadget} code as part
of a recipe for regulated star formation.  This approach leads to a
strongly hydrodynamically-coupled, multiphase ISM that does not
naturally produce galactic outflows.  To combat this the authors
added a ``superwind'' model where fluid elements in the star forming
region are ejected at fixed speed and are also hydrodynamically decoupled
until they leaves the galaxy.  \citet{Oppenheimer06} modified the model
in a manner referred to as the momentum-driven wind scenario so that the velocity of
the wind and the mass loading factor were related to the velocity
dispersion of the host galaxy.  The \textsc{gadget} code with 
superwind feedback prescriptions has been widely used in various problems such as damped
Lyman-$\alpha$ (DLA) absorbers \citep{Nagamine04a, Nagamine04b} and
the enrichment of the IGM at  
high and low redshifts \citep{Oppenheimer06, Oppenheimer09}.  According
to these works, superwind feedback is essential to suppress
overproduction of stars in galaxies and to reproduce the cosmic SFH at
high redshift. It also increases the local fraction of the warm-hot
intergalactic medium (WHIM) to a sufficient percentage (40\% to 50\%)
to account for the ``missing baryons'' at z = 0. Although aspects of
the model compare well with 
observations, some components do not.  For  
example, the feedback may eject a large amount of cool gas from the galactic
disks, which results in a low neutral hydrogen mass density
$\Omega_{\rm H I}$ at z $<$ 2 
\citep{Nagamine04a}. Also, the interaction between winds and the ISM
is usually not modeled in these simulations.   \citet{DallaVecchia08} found that the ISM
plays an important role in regulating the amount of wind that
escapes and the morphology of the galaxies. In their
model winds are not hydrodynamically decoupled, which naturally allows for variable mass loading.   

A refined version of energetic feedback is adiabatic feedback which
treats the overcooling problem by inhibiting gas from cooling until
the hot SN bubbles can be resolved \citep[e.g.,][]{Thacker00,   Kay02,
  SommerLarsen03, Stinson06}.  This is the approach used in this work. 
The pressure of the hot gas accelerates
the ISM to generate winds.  In the high resolution limit,
this method approaches direct ISM modeling.  Though energetic feedback is often referred
to as supernova feedback, it can be used to model several types of stellar
feedback such as winds and locally deposited radiation energy.  The
essential quantity is the energy injection rate as a function of the
mass in stars and the current age of the stellar population.  

\citet{Theuns02} used the \citet{Kay02} adiabatic feedback model that
turned off cooling for 10 Myrs for the feedback gas in their
cosmological simulations. They found enough
metals were carried by strong winds to produce C IV absorption lines
that agreed with observations.    \citet{Aguirre05} used the same
simulation to compare the optical depth of C IV and C III absorption
lines from simulations with observations. 

The properties of the enriched IGM are not only affected by winds but also
by gas cooling.  Winds can enrich galactic halos and the IGM so that
metal cooling significantly increases the cooling rates. 
\citet{Aguirre05} found that their simulated metal enriched gas was too
hot ($10^{5} \sim 10^{7}$ K) and suggested that a lack of metal cooling
was responsible for discrepancies between simulated and observed C IV
absorption.  \citet{Oppenheimer06} found better agreement when they
included the \citet{Sutherland93} metal cooling model.  The same model
was used by \citet{Choi09}, who investigated the effect of metal
cooling on galaxy growth and found that it increases the local star forming
efficiency and enhances accretion onto galaxies.  However,
\citet{Sutherland93} did not include photoionization due to a
ultraviolet (UV) radiation background which strongly affects the
ionization states of metal species and changes the cooling rates.
This is investigated in detail in the current work and also in 
\citet{Wiersma09a}.

Another important aspect of metal enrichment is the mixing of metals
between the wind and the surrounding gas.  The
interstellar medium is highly turbulent and SN explosions are likely
to be a major driver of the turbulence \citep{MacLow04}.  In addition,
large velocity shear (such as between a wind and a gaseous halo)
naturally generates turbulence and mixing.  
Turbulent mixing redistributes metals and thermal energy between the wind
fluid and the ambient gas.  This changes the metallicity,
temperature and future evolution of the gas. While metal mixing
  is expected in strong outflows, it is still unclear how mixing
  impact the IGM. For example, observations by \citet{Schaye07} found
  compact ($\sim$ 100 pc), transient C IV absorbers that are highly
  enriched, suggesting poor chemical mixing at small scales.  These
  absorbers were interpreted as enriched clumpy medium embedded within
  hot galactic wind fluids. If velocity shear is the major mechanism
  for turbulent mixing between winds and the surroundings,  then this
  poor mixing could be explained if the clouds are carried by hot
  winds at the same speed. However to investigate this in detail, one
  must resolve wind structures, which is beyond current cosmological simulations. In this work we will focus on subgrid turbulent mixing models in cosmological context.   In SPH simulations (which represent the majority of work in this area), the
fluid is modeled by discrete particles. This implies that newly injected 
metals are locked into specific particles.  For example, it was found
that the distribution of metals from SPH simulations is too inhomogeneous
compared with observations \citep{Aguirre05}. To assess the potential
importance of mixing, \citet{Wiersma09} 
used SPH-smoothed metallicities and compared it with conventional
particle metallicities, and found smoothing is able to generate significantly more
material with low metallicities.  This approach cannot capture
the spread of metals over time with its impact on cooling and the
thermal history of the gas.  
Directly modeling the turbulent ISM within a cosmological
simulation is far beyond current capabilities.  We
employed a variant of the \citet{Smagorinsky63} subgrid turbulent
diffusion model, in which unresolved turbulent mixing is treated as a
shear-dependent diffusion term. Metal cooling was calculated based on
the diffused metals so that its non-local effects could be investigated.  

In this work, we present an analysis of a series of SPH cosmological simulations
that incorporated adiabatic stellar feedback, detailed metal cooling
and turbulent mixing to study the evolution and enrichment of the IGM.
The feedback model was kept simple, following the adiabatic stellar feedback approach of \citet{Stinson06}.
This model has been calibrated via numerous galaxy formation studies \citep[e.g., ][]{Governato07}.
No additional wind prescriptions were used.  With this approach, outflows arise
from stellar feedback within the ISM and there is no
distinction between the feedback that regulates star formation
and that which drives galactic outflows.  Thus this work establishes a
baseline for the effectiveness of moderate stellar feedback coupled with
key physical process absent from other work to reproduce the
properties of the IGM.  These results may be compared with explicit wind models.
A further goal of this paper is to separately quantify the impact of
metal cooling and turbulent mixing on the SFH, the global properties
and the evolution of the IGM and its 
enrichment.  In this first paper of a series, we present general results.
The properties of specific metal absorbers in simulated quasar
spectra will be presented in a second paper.

This paper is organized as follows.  Section~\ref{method} describes
the models for cosmological hydrodynamics, star formation, supernova
feedback, metal cooling and metal diffusion.  Section~\ref{global}
examines the cosmic SFH, global H I fraction and Ly-$\alpha$ decrement in order to calibrate our models.
Section~\ref{mass_metal_evo} focuses on the evolution of the baryonic
mass, metal fractions and metallicities in stars and different gas
phases.  We compare those results to the observed metal fractions
and metallicities at different epochs, and with the simulations using
different subgrid feedback models from \citet{Oppenheimer06},
\citet{Dave07} and \citet{Wiersma09}.  In section~\ref{phasediagram}
we analyze the distribution of mass and 
metallicity in the density-temperature phase diagram at z = 0.  In section~\ref{wind} we
characterize our wind efficiency as a function of 
galaxy mass to obtain a better understanding of how
different phases of the IGM get enriched.  Where relevant,  we have
included detailed analysis of the effects of metal cooling and
diffusion, and comparisons with observations.  In the final 
section~\ref{summary}, we summarize and discuss the broader implications.

\section{Method and Simulations}
\label{method}

Table~\ref{list} lists the simulations used in this study. The    
cosmological parameters are ($\Omega_{m},\Omega_{\Lambda},
\Omega_{b}, h, \sigma_{8}$, n) = (0.279, 0.721, 0.0462, 0.701, 0.769,
1.0), consistent with the WMAP 5-year results \citep{Komatsu09}. The first three simulations in the table share the same initial
condition but have different subgrid models for metal cooling and
diffusion. The particle mass for gas is 
$m_{g} = 2.4 \times 10^{7} M_{\sun}$  and for dark matter is $m_{d} =
1.2 \times 10^{8} M_{\sun}$. When discussing he effects of metal cooling and diffusion, we will usually refer to the first simulation (mcd\_40\_256) "the reference run".  The last two
simulations were designed for a convergence study. They have 
the same initial condition but differ from the first three. The
high resolution case has particle masses  $m_{d} = 3.6 \times 10^{7}
M_{\sun}$ and $m_{g} = 4.5 \times 10^{6} M_{\sun}$. All the simulations have the
same star formation, metal production and SN feedback models described
below. The simulations were run to redshift zero, except for the
``mcd\_45\_512''case which was stopped at z = 2. 

\begin{table}
\caption{List of Simulations}
\begin{tabular}{@{}ccccc}
\hline
Name & Size (Mpc) & $N_{p}$ & Metal Cooling & Diffusion  \\ 
\hline
mcd\_40\_256 & 40.0 & $2 \times 256^{3}$ & Yes & Yes \\ 
nmc\_40\_256 & 40.0 & $2 \times 256^{3}$ & No  & Yes \\ 
nmd\_40\_256 & 40.0 & $2 \times 256^{3}$ & Yes & No \\ 
mcd\_45\_256 & 45.6 &  $2 \times 256^{3}$ & Yes & Yes \\ 
mcd\_45\_512 & 45.6 &   $2 \times 512^{3}$ & Yes & Yes \\ 

\hline
\end{tabular}
\label{list}
\end{table}

The simulations were evolved using the parallel SPH code \textsc{gasoline}      
\citep{Wadsley04}.  \textsc{gasoline} solves the equations of
hydrodynamics and includes radiative cooling.  Gravity is calculated
for each particle using a binary tree elements that span at most $\theta$ = 0.7
of the size of the tree element's distance from the particle. 

The star formation and feedback recipes used were the ``blastwave model''
described in detail in \citet{Stinson06}, and they may be summarized
as follows.  Gas      
particles must be dense ($n_{\rm min}=0.1\ cm^{-3}$) and cool ($T_{\rm max}$     
= 15,000 K) to form stars.  A subset of the particles that pass these           
criteria are stochastically selected to form stars based on the
commonly used star formation equation,             
                    
\begin{equation}                                                                
\frac{dM_{\star}}{dt} = c_{\star} \frac{M_{gas}}{t_{dyn}}                       
\end{equation}                                                                  
where $M_{\star}$ is mass of stars created, $c_{\star}$ is a constant star      
formation efficiency factor, $M_{gas}$ is the mass of gas creating the star,    
$dt$ is how often star formation is calculated (1 Myr in all of the             
simulations described in this paper) and $t_{dyn}$ is the gas dynamical         
time.  The constant parameter, $c_{\star}$ was set to 0.05 so that          
a simulated isolated model Milky Way matches the        
\citet{Kennicut98} Schmidt Law \citep{Stinson06}. Gas particles passing all the criteria form one star.  The mass of star formed remains constant throughout the simulation and is set as one-third of the original  gas mass ($8.0 \times 10^{6}$ M$_\odot$ for the 40 Mpc, $256^{3}$ runs and $1.5 \times 10^{6}$ M$_\odot$ for the $512^{3}$ run). 

At the resolution of these simulations, each star particle represents
a large group of stars.  Thus, each
particle represents a stellar population covering the entire initial
mass function presented in \citet{Kroupa93}.  Star masses are
converted to stellar lifetimes as described in \citet{Raiteri96}.
  We implemented feedback from SN II, SN Ia and stellar winds. Metal enrichment from SN II and SN Ia follows the model
  of \citet{Raiteri96}, but metal production of AGB stars is not included.
Stars larger than 8 $M_\odot$ explode as SN II during the
timestep that overlaps their stellar lifetime after their birth time.
The explosion of these stars is treated using the analytic model for
blastwaves presented in \citet{McKee77} as described in detail in
\citet{Stinson06}.  Gas cooling is suppressed during the expansion of
the SN bubble.  While the blast radius and the cooling shutoff time
are calculated using the full energy output of the supernova, less
than half of that energy is transferred to the surrounding ISM,
$E_{SN}=4\times10^{50}$ ergs.  The rest of the supernova energy is
assumed to be radiated away.  The energy is transferred to the ISM by
volume weighting (as used in \citet{Mashchenko08}).  Each affected
gas particle, with mass $m_{i}$ and density $\rho_{i}$ receives a
fraction of the SN energy and metals proportional to $m_{i}W_{ij}/\rho_{i}$,
where $W_{ij}$ is the SPH smoothing kernel.  A similar method was also
adopted in \citet{Wiersma09} for the distribution of metals and the
energy from SN Ia.  The Blastwave model, as currently 
implemented, has a bias for more cooling suppression at earlier epochs
relative to other approaches.  Numerical feedback recipes differ
substantially from code to code and their results vary with
resolution.  This provided motivation for the characterization of star
formation regulation and wind generation in these simulations that
is presented in section~\ref{wind}.
                   
For SN II, metals produced in stars are released as the main sequence
  progenitors die and distributed to the same gas within the blast
  radius as is the supernova energy ejected from SN II. Iron and
  Oxygen are produced in SN II according to the analytic fits used in
  \citet{Raiteri96} using the yields from \citet{Woosley95}:                                                               
                                                                                
\begin{equation}                                                                
M_{Fe} = 2.802 \times 10^{-4} M_\star^{1.864}                                   
\end{equation}                                                                  
                                                                                
\begin{equation}                                                                
M_{O} = 4.586 \times 10^{-4} M_\star^{2.721}                                    
\end{equation}                                                                  
                                                                                
Feedback from SN Ia also follows the \citet{Raiteri96} model, as
described in detail in \citet{Stinson06}. Radiative cooling was not
disabled for SN Ia. Each SN Ia produces 0.63 $M_\odot$ Iron and 0.13
$M_\odot$ Oxygen \citep{Thielemann86} and the metals are ejected into
the nearest gas particle for SN Ia. Stellar wind feedback was
implemented based on \citet{Kennicutt94}, and the returned mass
fraction was determined using a function derived by
\citet{Weidemann87}. The returned gas has the same metallicity as the
star particle.   
                                           
\subsection{Metal Cooling under the UV Radiation}
\label{cooling}

Collisionally excited metal ion species can significantly enhance
the cooling of the IGM by their line transitions (see e.g.,
\citet{Sutherland93}). 
Metal cooling thus affects the gas temperature, the ionization 
states of the observable metal species and ultimately the dynamics.  It was not feasible to calculate 
metal cooling rates during the simulation.  Instead, tabulated
cooling rates were interpolated during the simulations as in
\citet{Oppenheimer06}, \citet{Wiersma09} and \citet{Choi09}.  Metal cooling  
rates were determined from the density, temperature,
metallicity of the gas and the radiation background.  While the IGM and ISM are
exposed to UV radiation, widely-used
metal cooling rates  \citep[e.g., ][]{Sutherland93, Gnat07} were
usually calculated assuming collisional 
ionization equilibrium (CIE), which is only
valid when the radiation 
background is absent. A recent study by \citet{Wiersma09a} has also investigated
the effects of radiation on cooling in the temperature range $10^{4} <
T < 10^{8}$ and also shows that UV largely suppresses the cooling rates and
shifts the peaks of the cooling curves to higher temperatures.             

The radiative cooling was separated into three components: 
\begin{equation}
\Lambda = \Lambda_{H, He} + \Lambda_{metal} + \Lambda_{Comp}
\end{equation}
where $\Lambda_{H, He}$ is net cooling due to primordial species (H,
H$^{+}$, He, He$^{+}$ and He$^{++}$), $\Lambda_{metal}$ is the rate due to
metals, and $\Lambda_{Comp}$ is the Compton cooling/heating.  For the
primordial gas the ionization, cooling and heating 
rates are calculated directly from the ionization equations with rates matching \citet{Abel97} .
This enables our simulations to capture the
non-equilibrium cooling of primordial species.  Although our study
focuses mainly on epochs where the IGM is mostly in ionization
equilibrium, non-equilibrium effects are important at earlier
redshifts, especially during reionization, where temperature
changes of several thousand degrees occur as a result of
non-equilibrium heating. 

We adapted the relation from \citet{Jimenez} to estimate the helium
mass abundance $Y$ as it varies with metallicity. For metal mass abundance $Z<0.1$, we assume $Y = Y_{p} +
(\Delta Y /\Delta Z)Z$, where $Y_{p} = 0.236 $ is the helium abundance for
primordial gas, and $\Delta Y / \Delta Z = 2.1$ is the ratio of helium
mass to metal mass produced in stars.   If the metal mass abundance
exceeded 0.1 ($\sim$ 10$Z_{\sun}$), $Y$ is linearly decreased so that
when $Z=1.0$ (100\% metals), $Y=0$.           

The cooling and heating rates due to metals under the ultraviolet(UV)
radiation background were calculated using the
photoionization code \textsc{cloudy} (version 07.02 \citet{Ferland98}). 
\textsc{Cloudy} output assumes that metals are in ionization equilibrium, a good approximation
when extragalactic UV radiation is present.  The UV radiation
field generally varies its function form in space and time. We assume
a uniform background obtained from extracting the \textsc{cloudy}
(07.02) built-in extragalactic UV 
background which includes radiation from both quasars and galaxies \citep{HM05}.The radiation field is a function of wavelength, and it turns on at redshift z = 8.9 and evolves with
redshift to z=0. The same UV background was also adopted to calculate the cooling due to primordial species. 

Metal cooling depends on redshift because the radiation background
evolves with time. Although our simulations traced the formation of
alpha 
elements and iron separately, our metal cooling assumed relative
solar abundances for simplicity.  We used the \textsc{cloudy} default solar
composition, which contains the first thirty elements in the
periodic table (Data compiled from \citet{GS98, H01, AP01, AP02}, see
Table 9 of the \textsc{cloudy} version 07.02 documentation for details). The
metal cooling rates were calculated by subtracting the primordial
cooling rates from the cooling rates of the total 30 elements,
including H and He. We found that the net metal cooling rate is linear
with metallicity, which allows us
to tabulate the rates for solar abundance only and scale with
metallicity (equation~\ref{coolingeq}). The metal cooling rates can
thus be written as:

\begin{equation}
\label{coolingeq}
\Lambda_{metal}(T, \rho, z, Z) = \frac{Z}{Z_{\sun}}\Lambda_{metal,\sun}(T,\rho, z) 
\end{equation}

where T, $\rho$, z and Z are the temperature, density, redshift and
metallicity of the gas, respectively. 

The metal cooling rates were tabulated in log(T) (in unit of K, from
100 K to $10^{9}$K with $\Delta$log(T)= 0.05), log($n_{H}$) (where
$n_{H}$ is the hydrogen number density, from $10^{-9}$ cm$^{-3}$ to
$10^{3}$ cm$^{-3}$ with $\Delta$log($n_{H}$) 0.1), and redshift z (8.9
$\lid$ z $\lid$ 0 with grid size 0.1).  The cooling and heating rates
were interpolated separately in log space in the simulation. The RMS
of relative interpolation errors for cooling and heating are less than
1 percent.   We assumed the gas was optically thin to the ionizing
background for all wavelengths, which is valid for the IGM.  We included
metal cooling down to 100 K which can occur in dense environments or
in the IGM that is cooled by adiabatic expansion, although the latter
is generally not not enriched. At
low temperature we turned off the molecular hydrogen cooling, because
$H_{2}$ in the IGM is not in ionization equilibrium and
\textsc{cloudy} gave unrealistically high $H_{2}$ cooling rates by
assuming equilibrium (also see \citet{Smith08}). Beyond $z>8.9$
cooling was calculated with effectively no UV background.

\begin{figure}
\includegraphics[width =\columnwidth]{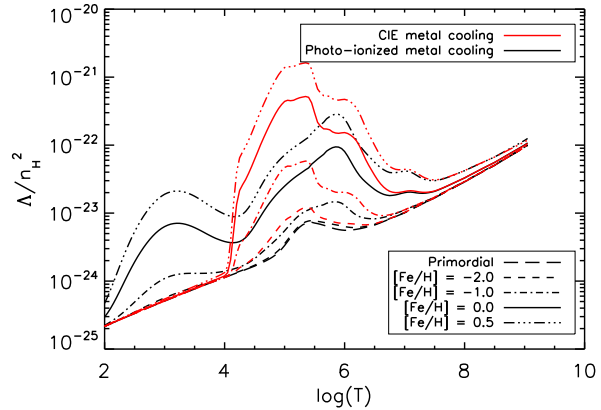}
\caption{Cooling rates normalized by $n_{H}^{2}$ (in erg s$^{-1}$
  cm$^{3}$) with metallicities 
  [Fe/H] = -2.0, -1.0, 0 and 0.5.  The gas has density $n_{H} =
  10^{-5} $ cm$^{-3}$. {\it Black}: total cooling rates (include H, He and
  metals) under the z=3 extragalactic radiation background from
  \citet{HM05};  {\it Red}: Sum of cooling rates of H and  He
  under the same   radiation background,  and metal cooling rates, 
  calculated without UV background}  
\label{figcooling}
\end{figure}

Figure~\ref{figcooling} shows cooling rates normalized by $n_{H}^{2}$
as a function of temperature at various metallicities. The black
curves show the total cooling 
rates with z=3 radiation background, while the red curves show the CIE
metal cooling (no UV) plus the cooling of H and He under the UV
background, a model often adopted in cosmological simulations.
With or without UV, the presence of metals can increase
cooling rates by up to several orders of magnitude. In the
temperature range $10^{4}< T < 10^{8}$ K, radiation ionizes the plasma
and reduces the number of ions that can be
collisionally excited therefore decreases the cooling rates
significantly (e.g., about an order of magnitude at $10^{5}$ K in Figure
\ref{figcooling}).  The cooling peak also shifts from $10^{5} K$ to
$10^{6}K$ due to the change of ionization states of the metal
elements.  From 100 K to
$10^{4}$ K, the UV background 
increases the number density of free electrons ($n_{e}$) hence
enhancing forbidden line cooling from low ionized species such as C
I, C II, Si I, Si II and O I.  Forbidden line cooling is the dominant
cooling process in enriched IGM at low temperatures.  
At low densities ($n_{H} \la 10^{-4}
cm^{-3}$), photo-ionization also 
creates some higher-ionized species with strong magnetic dipole
transitions (e.g. Ne V and Ne VI).  These ions can cool the plasma more
efficiently, producing a local cooling peak around $10^{3}$K (Figure
\ref{figcooling}).  However with an increase in density these ions
disappear and the major coolants are the less ionized species mentioned,
consistent with ISM cooling models \citep[e.g.][]{Wolfire03}. Overall,
the radiation background increases the metal cooling rate
substantially between 100 K and $10^{4}$ K over CIE.  Figure~\ref{figcooling} 
shows unequivocally that simply adding the CIE model of metal cooling on the
primordial cooling rates results in dramatic over or under estimation of the
cooling rates, depending on the temperature. The amplitude of the UV
background used here is however too high, as implied in the Lyman
$\alpha$ flux decrement evolution in Section \ref{flux_decrement}.
Thus the effect of photo-ionization on metal cooling rate may be
overestimated.

\subsection{Turbulent Metal Diffusion}
\label{diffusion}

As a Lagrangian particle method, SPH does not include any implicit
diffusion of scalar quantities such as metals.  \citet{Wadsley08}
demonstrated that this has physically incorrect consequences for even
simple processes such as convection and Rayleigh-Taylor instabilities.
Including a simple model for turbulent mixing (using a diffusion
coefficient, $D = C\,\Delta\,v\ h_{\rm SPH}$ based on the pairwise
velocity, $\Delta\,v$, at the resolution scale, $h_{\rm SPH}$ and $C
\sim 0.1$) \citet{Wadsley08} were able to match Eulerian grid code
results (which 
must mix due to the necessary advection estimates).  In particular, it
became possible to generate similar non-radiative galaxy cluster
entropy profiles with SPH as with high resolution grid codes.  This
was a major discrepancy in the \citet{Frenk99} cluster
comparison. \cite{Greif09} implemented a similar scheme
and applied it to simulating supernova remnants.  As discussed
previously, galactic outflows should be highly turbulent and thus
mixing is essential for IGM studies. 

Turbulent mixing models have a long history in environmental and
engineering fluid mechanics.  \textsc{gasoline} now incorporates a more
robust mixing estimator similar to that first proposed by
\citet{Smagorinsky63} for the atmospheric boundary layer, 
\begin{eqnarray}
\frac{dA}{dt}|_{\rm Diff} & = & \nabla (D \nabla A), \nonumber \\
D & = & C\,|S_{ij}|\,h^2,
\end{eqnarray}
where $A$ is any scalar, we use the trace-free shear tensor for
$S_{ij}$ and $h$ is the measurement scale (here $\sim h_{SPH}$).  This
choice for $S_{ij}$ results in no diffusion for compressive or purely
rotating flows.  In SPH terms the diffusion expression for a scalar
$A_p$ on particle $p$ is computed as follows,
\begin{eqnarray}
\tilde{S}_{ij}|_p &=& \frac{1}{\rho_p}\sum_q m_q (v_j|_q-v_j|_p)
\nabla_{p,i} W_{pq}, \nonumber \\ S_{ij}|_p &=& \frac{1}{2}
(\tilde{S}_{ij}|_p+\tilde{S}_{ji}|_p) - \delta_{ij} \frac{1}{3}\ {\rm
Trace}\ \tilde{S}|_p, \nonumber \\ D_p &=& C\ |S_{ij}|_p|\ h_p^2, \\
\frac{dA_p}{dt}|_{\rm Diff} &=& -\sum_q m_q
\frac{(D_p+D_q)(A_p-A_q)(\mathbf{r}_{pq}\cdot\nabla_p
W_{pq})}{\frac{1}{2}(\rho_p+\rho_q)\,\mathbf{r}_{pq}^2}, \nonumber
\end{eqnarray}
where the sums are over SPH neighbours, $q$, $\delta_{ij}$ is the
Kronecker delta, $W$ is the SPH kernel function, $\rho_q$ is the
density, $\mathbf{r}_{pq}$ is the vector separation between particles
$v_i|_q$ is the particle velocity component in direction $i$,
  $\nabla_p$ is the gradient operator for particle p (operating on the
  SPH kernel function) and $\nabla_{p,i}$ is the {\it i}th component
  of the resultant vector. The difference between this model and the
ones used in \citet{Wadsley08} and \citet{Greif09} is that the
diffusion coefficient is calculated according to a turbulent mixing model
instead of simply using velocity differences. The coefficient depends on
the velocity shear hence it better models the mixing in shearing
flows. On the other hand, if there is no shearing motion between two
phases of fluids (such as a clumpy medium embedded in hot wind fluid
and the two move with same speed), then no turbulent diffusion is
added. The Smagorinsky model has also been used successfully in other
field such as weather modeling. 

A coefficient value of order $0.05-0.1$ is expected from turbulence
theory (depending on the effective measurement scale, $h$).  It was
found that a conservative choice of $C=0.05$ was sufficient to match
the cluster comparison.  This diffusion was applied to thermal
  energy and metals in all runs except ``nmd\_40\_256''. Diffusion of
  density results naturally from the noise in the velocity field which
  moves particles around. 

\section{Global Properties of the Simulation}
\label{global}
Before embarking on a detailed examination of the metal distribution and evolution in the
IGM, it is worth establishing the basic properties of the
simulated volume with respect to the star formation history (SFH), the Lyman-$\alpha$
forest and the evolution of total neutral hydrogen.  These
properties directly reflect the effectiveness of the feedback
processes and allow the approach to be calibrated with respect to observations.

\subsection{Star Formation History} 
\label{sfr}
\begin{figure}
\includegraphics[width =\columnwidth]{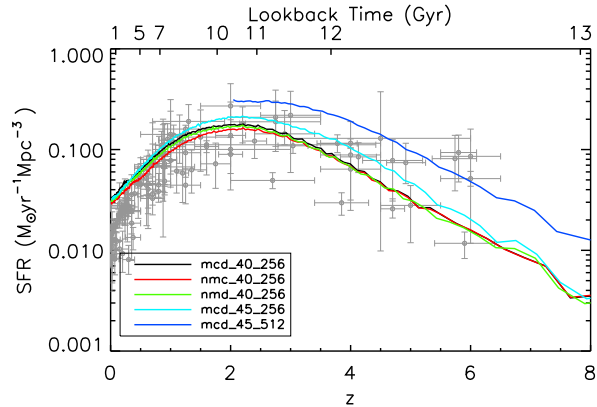}
\caption{Star formation rate (SFR) density (in unit of M$_{\sun}$
  yr$^{-1}$ Mpc$^{-3}$) as a function of 
  redshift.  Data were compiled from \citet{Hopkins04} and scaled to
  the same IMF and cosmology as in our simulations. The
    moderate resolution simulations give SFR consistent with
  observations down to redshift 0.5. The high resolution run has
  significantly higher SFR but still in broad agreement with
  observations.} 
\label{figsfr}
\end{figure}

Figure~\ref{figsfr} shows the evolution of global star formation rate (SFR)
as a function of redshift.  The observational data with error bars were
adapted from \citet{Hopkins04} and scaled to the same IMF and
  cosmology as used in our simulations. The figure shows that all simulations
with moderate resolution (with or without metal cooling and
diffusion) produce the SFH that is consistent with observations for
$z \ga 0.5$. 
The peak of star formation lies around $z \sim 2$ for all simulations. 
Without superwind models, \citet{SH03} and \citet{Oppenheimer06} found 
higher SFR than the observations and the SFH is peaked at
$z\sim 4$. Our results suggest that the suppression of SFR can
  also be attained using feedback models without explicit winds.
Until $z \sim 0.5$, our 256$^3$ simulations show no overproduction of
stars while matching other observables.  
Note that while the 512$^3$ does show high SFR, it was a direct
convergence test with all parameters fixed.  The parameters
affecting SF efficiency should be adjusted downwards with
increasing resolution for consistent outcomes at $z=0$ \citep[c.f.][]{Stinson06}.

The simulations do form too many stars at $z < 0.5$. One possible
reason is that AGN feedback becomes more relevant because large galaxies or  
clusters become more prevalent at lower redshift, and the supermassive black
holes in these objects can generate strong AGN feedback effects. Since
AGN were not included in our model, the global SFR is higher than
observations.            

 The simulation without metal cooling
produced fewer stars at $z<5$. The difference in SFR between the
simulations with and without metal cooling increases from $\sim$ 5 \%
at z =4 to $\sim$ 20 \% at z=1, 
then decreases to 10 \% at z=0. Metal cooling enhances the cooling of
the IGM and therefore  its accretion rate onto
galaxies. \citet{Choi09} also found enhancement  
of SFR with metal cooling, though their SFR was generally enhanced by
20\% to 30\% over all redshifts below z=15, and in some runs by 50\%
at z=1. The discrepancy may result from their adoption of
\citet{Sutherland93} metal cooling rates that are significantly 
higher than what we used, since it does not include the radiation
background. However, \citet{Schaye10} used the cooling model from
  \citet{Wiersma09a} which also includes the suppression of metal
  cooling by UV background,  yet still found a larger enhancement of
  SFR comparable to \citet{Choi09}. This is possibly due to the higher
  metallicities in stars and the ISM in their simulations
  \citep{Wiersma09}.  In particular, they included mass loss by
  AGB stars which may largely contribute to the enrichment of the
  ISM.

The simulation without metal diffusion (but with metal cooling) also
produced slightly fewer stars (about 6 \% mean) than the reference
run at $z<4$,  because metal diffusion allows enriched particles to mix
their metals, so more particles experience metal cooling and turn into
stars. On the other hand, the metallicities of the original enriched
particles are reduced 
because of the diffusion, thus those particles have less strong
cooling hence are less likely to form 
stars. Our result suggests that the first factor dominates.   Since
the effect of metal diffusion relies heavily on metal cooling, it
is smaller than in the case where the metal cooling was turned
off. 

The effect of resolution on the SFH is shown in the convergence runs
  ``mcd\_45\_256''(cyan) and ``mcd\_45\_512'' (blue). At z = 8, the
  high resolution run produces 3 times larger SFR. Although the
  result does not converge by z = 2, the difference between these two
  runs does steadily decrease with time. At z = 2, it reaches about
  50\%. The SFH from ``mcd\_45\_512'' is close to the high
  end of the observational data but still consistent with them. 
  The increased star formation is the same magnitude seen in the convergence tests in    
\citet{Stinson06}, where it was determined that higher resolution simulations        
produce more high density gas.  High density gas forms more stars either because it    
surpasses the density threshold where gas in the low resolution simulations does       
not, or because star formation is a function of gas density in the Kennicutt-Schmidt law.  Since the difference is most prominent in the early epochs of our            
simulations when galaxies are small, it is likely that many dwarf galaxies surpass     
the gas density threshold in the high resolution that do not in the low resolution     
simulations. 
 % The SFR increment is also seen in the convergence test in
  % \citet{Stinson06} where isolated galaxies (at same z) were simulated with
  % 10-100 times higher mass resolutions. As discussed in
  % \citet{Stinson06}, the reason the probable explanation is that more gas
  % particles can reache higher density in a higher resolution
  % simulation. Also, the difference is most prominent at early epochs, which
  % is probably because at higher redshift  galactic halos are smaller and more likely
  % to be underresolved when the resolution is low, and hence the SFR in
  % these objects could be underestimated \citet{Schaye10}.  
  The effect of initial condition variation is also worth noting: the
  ``mcd\_45\_256'' run has slightly lower (but comparable) resolution
  than the reference run. However, it has a higher SFR at z $>$ 1.

\subsection{Evolution of $\Omega_{\rm H I}$} 
\label{sec_omegaHI}
\begin{figure}
\includegraphics[width =\columnwidth] {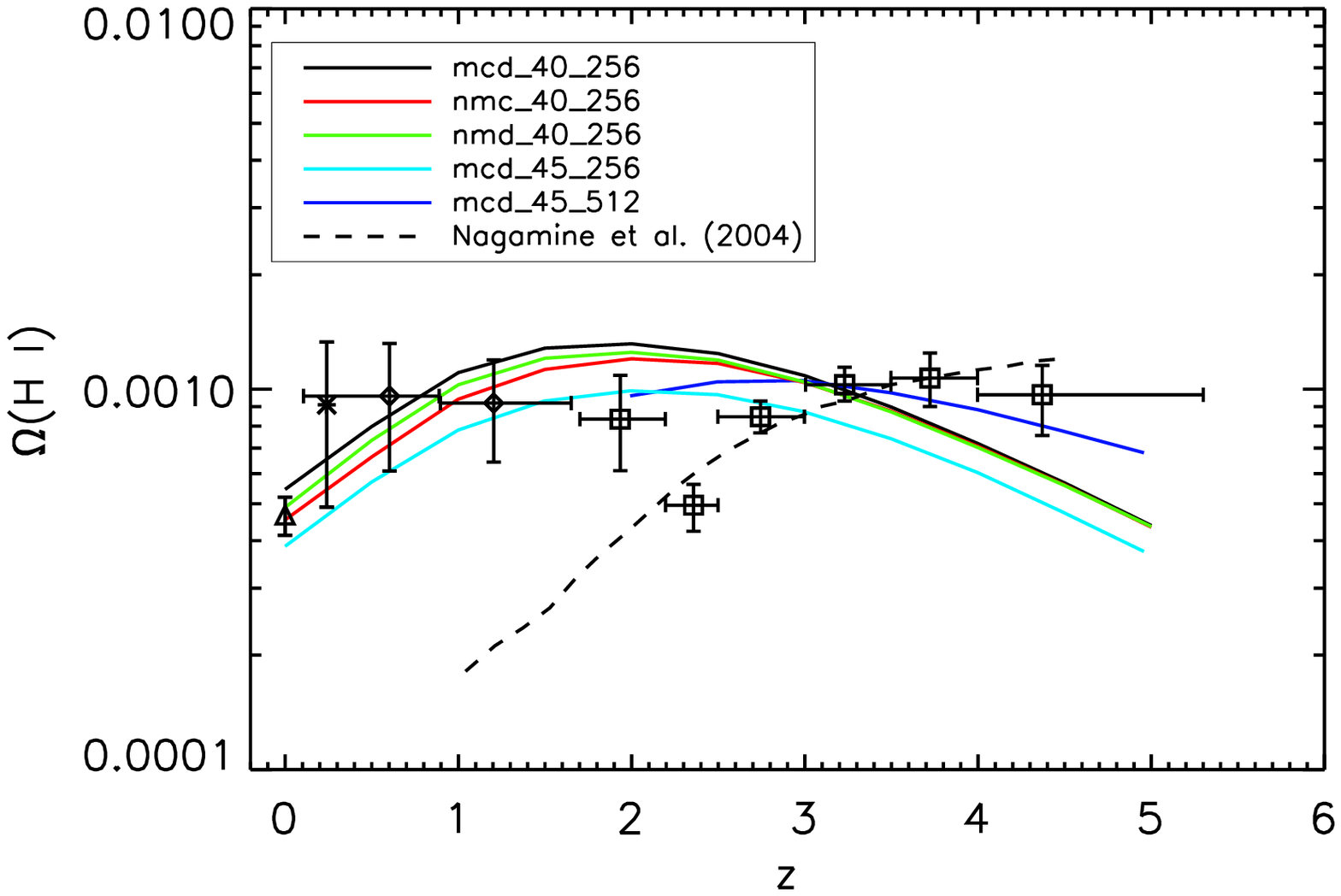}
\caption{Evolution of total neutral hydrogen density in units of
  critical density today ($\Omega_{\rm H I}$). Observational data points:  {\it
    Square}: \citet{Prochaska05};  {\it Diamond}: \citet{Rao06}; {\it
    Asterisk}: \citep{Lah07}; {\it Triangle}: \citet{Zwaan05}. } 
\label{OmegaHI}
\end{figure}

The evolution of the cosmic mass density in neutral hydrogen
($\Omega_{\rm H I}$)with 
redshift and its relation to stellar mass density is one of the key observables 
for understanding the
interaction between gas and galaxies. Observations of damped
Ly-$\alpha$ systems (DLAs) \citep{Prochaska05, Rao06, Prochaska09} and the
H I 21-cm emissions \citep{Zwaan05, Lah07} (shown in Figure
\ref{OmegaHI}) suggest that $\Omega_{\rm H I}$ does not evolve substantially
 through cosmic time,  even though the stellar
mass density keeps increasing. This implies that there is a steady supply of
gas cooling onto galaxies, providing fuel for star formation. The
decrease of H I at z $ \sim $ 2.3 has been linked to violent 
feedback processes, including SN, galactic winds and AGN activities
\citep{Wolfe05, Prochaska09}, but it is not clear why the amount of H I
increases back at z $<$ 2.3. \citet{Prochaska09} argued that the data at
z $\la$ 2 was biased high and $\Omega_{\rm H I}$ should remain constant from
z = 2.2 to z=0 (the last 10 Gyrs).  The non-evolution of $\Omega_{\rm H I}$
indicates that star formation is self-regulated such that gas
accretion, star formation and feedback processes balance each other.
Numerical simulations without mass loss through winds \citep{Cen03}
found $\Omega_{\rm H I}$ is several times higher. \citet{Nagamine04a}
included the superwind model from \citet{SH03} in their simulations so
that neutral gas could be ejected  from the galaxies, which produced
the correct amount of $\Omega_{\rm 
  H I}$ at $z>2$, but at $z<2$ too much material was blown away,
resulting in a deficit of $\Omega_{\rm H I}$ (dash line in
Figure~\ref{OmegaHI}).   

Figure~\ref{OmegaHI} shows the evolution of $\Omega_{\rm H I}$ in our 
simulations.  Because most neutral
hydrogen resides in damped Ly-$\alpha$ systems, which are clouds with H
I column densities larger than $2 \times 10^{20} $ cm $^{-2}$
\citep[][and references therein]{Wolfe05}, the gas is mostly
self-shielded from external ionizing photons. Self-shielding
was not modeled during the simulations,  however we used a radiative
transfer post-processor to recalculate the ionization states
of hydrogen as described in \citet{Pontzen08}. The
ionizing background was the standard
\citet{HM05} reduced by a factor of 2 (as discussed in section~\ref{flux_decrement} below).
However, variations of this magnitude have
practically no impact on the dense gas that dominates $\Omega_{\rm H I}$. 

Figure~\ref{OmegaHI} shows that after we applied the self-shielding
correction, our feedback model produced similar $\Omega_{\rm H I}$ to
observations from $z \sim 3.5$ down to $z=0$ for the runs with
moderate resolutions. At higher redshift these simulations underestimate $\Omega_{\rm H I}$, possibly because of insufficient resolution (discussed below).The result indicates that our
feedback models effectively moderate star formation but not too
strongly, so that gas accretion onto disks is not disrupted. Thus it
can maintain the steady supply of neutral hydrogen to galactic
disks. The shape of the $\Omega_{\rm H I}$ relation follows the SFH, which
reflects the relation between the H I density and SFR density,
i.e. the Schmidt Law for star formation.  The
decrement at $z = 2.3$ is not reproduced in our simulations,
indicating that more violent feedback maybe necessary at this limited
redshift range if this feature is confirmed.  
We will investigate wind models in high resolution
single galaxy and galaxy groups  in future work.   

As expected, Figure~\ref{OmegaHI} shows that  metal cooling and metal
diffusion increases the amount of 
neutral hydrogen. The metal cooling effect is obvious at  z $\la$ 3
and  at z = 0 the increase is 
$\sim$ 17 \%. Metal diffusion has a smaller effect.  The reasons for
these effects are similar to the ones discussed in SFR analysis in
section~\ref{sfr}.  

With eight times higher resolution, the simulation ``mcd\_45\_512''
  produces a significantly higher $\Omega_{\rm H I}$ and a flatter
  curve at z $>$ 2.5 which make the result compare better with observations. At z = 5, the high resolution run contains about 80\% more neutral hydrogen.  At z =  2 the $\Omega_{\rm H I}$ curve joins the moderate resolution run  ``mcd\_45\_256'' but has a slightly steeper decreasing slope towards lower z.  The difference in $\Omega_{\rm H I}$ is likely because the high resolution run better resolves small halos, thus increasing the amount of self-shielded gas.  Since small halos dominate at high redshift,  the difference is most obvious there. We note that as a result of cosmic variance the ``mcd\_45\_256'' simulation contains 15-30\% less HI throughout  the simulation than the fiducial run even though it has a higher SFR.
 %It is also interesting to note the variance caused by initial condition.  Comparing with the standard run, the ``mcd\_45\_256'' simulation contains less (15\% to 30 \%) neutral hydrogen through cosmic history,  although it has a larger SFR (Figure \ref{figsfr}).  
 Thus the relationship between SFR and the gas reservoir $\Omega_{\rm H I}$ is not always monotonic as seen in the cases for ``nmc\_40\_256'' and ``nmd\_40\_256''.  Higher SFR consumes more neutral gas thus may decrease the amount of $\Omega_{\rm H I}$.

\subsection{Lyman-$\alpha$ Flux Decrement}
\label{flux_decrement}
 
The Lyman-$\alpha$ forest traces the neutral hydrogen in the diffuse
IGM. Since the IGM is optically thin, the mean flux decrement of the
Lyman-$\alpha$ forest, D$_{\rm Ly-\alpha}$,  is sensitive to the UV
radiation and can be used to calibrate the UV background in
simulations.  We generated the $Ly-\alpha$ spectra by tracing a lines
of sight through every simulation box. Each line of sight consists 25
lines segments, which were spread 
through the entire box such that the periodicity of the box was used
to eliminate any discontinuity in the resulting spectra. Particles
that lay within two smoothing lengths measured perpendicularly to their
closest line segments contributed to the 
absorption along the line,  and  the amount of contribution was
weighted using the SPH smoothing kernel. The particles' peculiar
velocity, thermal velocity and Hubble expansion were all taken into
account.  The H I, He I and He II fractions were calculated during the
simulation as described in Section~\ref{cooling}. The Ly-$\alpha$
optical depth at each velocity due to Doppler broadening was calculated
and then convolved with the Lorentz profile.  

\begin{figure}
\includegraphics[width=\columnwidth]{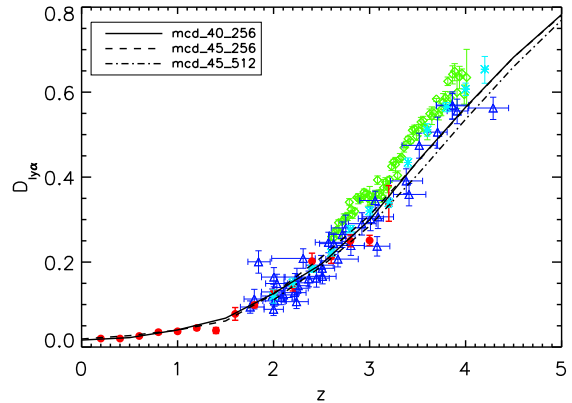}
%\caption{The evolution of the mean flux decrement of the Ly-$\alpha$ forest from our 
  %reference simulation.  {\it Solid line}  This work, with the UV
  %background attenuated by a factor of 2. {\it Dashed line} measurements
  %from \citep{Press93}. {\it Square symbol}: measurements from
  %\citet{Rauch97}.  {\it Triangle symbol with error bars}:
  %observations data from \citet{Kirkman07}.   }
\caption{The evolution of the mean flux decrement of the Ly-$\alpha$ forest from our reference simulation (``mcd\_40\_256'') and the convergence test (``mcd\_45\_256'' and ``mcd\_45\_512'').  Color symbols with error bars indicate the observational data.  {\it Green diamond}: \citet{Bernardi03}, data here are derived from their original evolution of effective optical depth, $\tau_{\rm eff}$ in their Fig. 4;  {\it Blue triangle}: \citet{Schaye03}, data here are derived from the original $\tau_{\rm eff}$ after removal of pixels contaminated by metal lines (Table 5);  {\it Red solid dot}: \citet{Kirkman07};  {\it Cyan asterisk}: \citet{Faucher-Giguere08}, derived from their $\tau_{\rm eff}$ evolution after contamination correction (Table 1). }
\label{dlyal}
\end{figure}

Figure~\ref{dlyal} depicts the evolution of $D_{Ly\alpha}$ compared to
the observations of \citet{Bernardi03}, \citet{Schaye03}, \citet{Kirkman07} and \citet{Faucher-Giguere08}.  When we attenuate the UV spectrum by factor of 2,
our results agree with most observations up to z $\sim 3.5 $. At
higher redshift, our results are slightly lower than, but still
broadly consistent with the observational data. The biggest difference
is seen when comparing to \citet{Bernardi03}. The discrepancy may
indicate that more UV attenuation for z $\ga$ 3 in our simulation is
necessary.  However, observations of the Ly$\alpha$ forest at high
redshift may be subject to large uncertainties due to the continuum
correction \citep{Bernardi03} and metal line contamination
\citep{Schaye03}.  The \citet{Schaye03} and \citet{Faucher-Giguere08}
data are contamination corrected and our result compare better with
these data. Hence, a factor of 2 seems sufficient and we use this
value for all subsequent calculations. This attenuated UV background
is only used in the analysis, but not in the entire
simulation. Although for the diffuse, highly ionized primordial
  IGM, the effect of photo-ionization on gas dynamics is small
  \citep{Croft98}, UV does affect metal cooling and hence star
  formation. Thus, by using the original UV background without
  attenuation, our metal cooling calculation may have overestimated
  the effect of photo-ionization.  However, along with other workers,
  we are also neglecting local ionizing sources which correct the UV
  upward to a significantly greater extent within galaxies.  These
  details of ISM physics are potentially important but beyond 
  current capabilities where we must rely upon relatively crude ISM models.

Focusing on the convergence test, with high resolution (dot-dashed line) the Ly$\alpha$ decrement evolves very similarly with redshift,  but is a few percent smaller than the lower resolution (dashed line) run.  The decrease in $D_{Ly\alpha}$  is consistent with the increase in SFR and $\Omega_{\rm HI}$. It can be interpreted as when more gas accretes onto galaxies and forms stars for high resolution, the H I content in the IGM (probed by the Ly$\alpha$ forest) decreases.

%\section{Distribution Gas and Metals} 
\section{Evolution of Gas, Metal Fractions and Metallicity}
\label{mass_metal_evo}

We now examine the enrichment of the IGM through 
the metal distribution in different gas phases.
We examine how they evolve and the effect of the metal
diffusion and cooling.  The gas phases were defined using the convention
of \citet{Wiersma09}.  All the gas that has hydrogen number
densities $n_{H} > 0.1$ cm$^{-3}$ is star forming gas (SF gas). Of all
the non-star forming gas (non-SF gas),  gas with $ 10^{5}K < T <
10^{7} K$ is the warm-hot intergalactic medium (WHIM) and gas 
with $T > 10^{7} K$ is the intracluster medium (ICM). Of the
cooler medium ($T < 10^{5} K$), that with overdensity
$\rho/\rho_{mean} > 100 $ is associated with galactic halos,
while that with $\rho/\rho_{mean} < 100 $ is the diffuse IGM.  Note
that these definitions are somewhat arbitrary and 
there is no well-defined border for each gas phase. In particular, at low redshift substantial part of the WHIM comes from the shock-heated gas accreting onto galaxies so it is likely to be associated with galactic halos. However, by this definition gas is roughly divided  according to its location and key physical
processes.   

\subsection{Evolution of Gas and Metal Fractions}
\label{sec_fractions}

\begin{figure*}
\includegraphics[width=175mm]{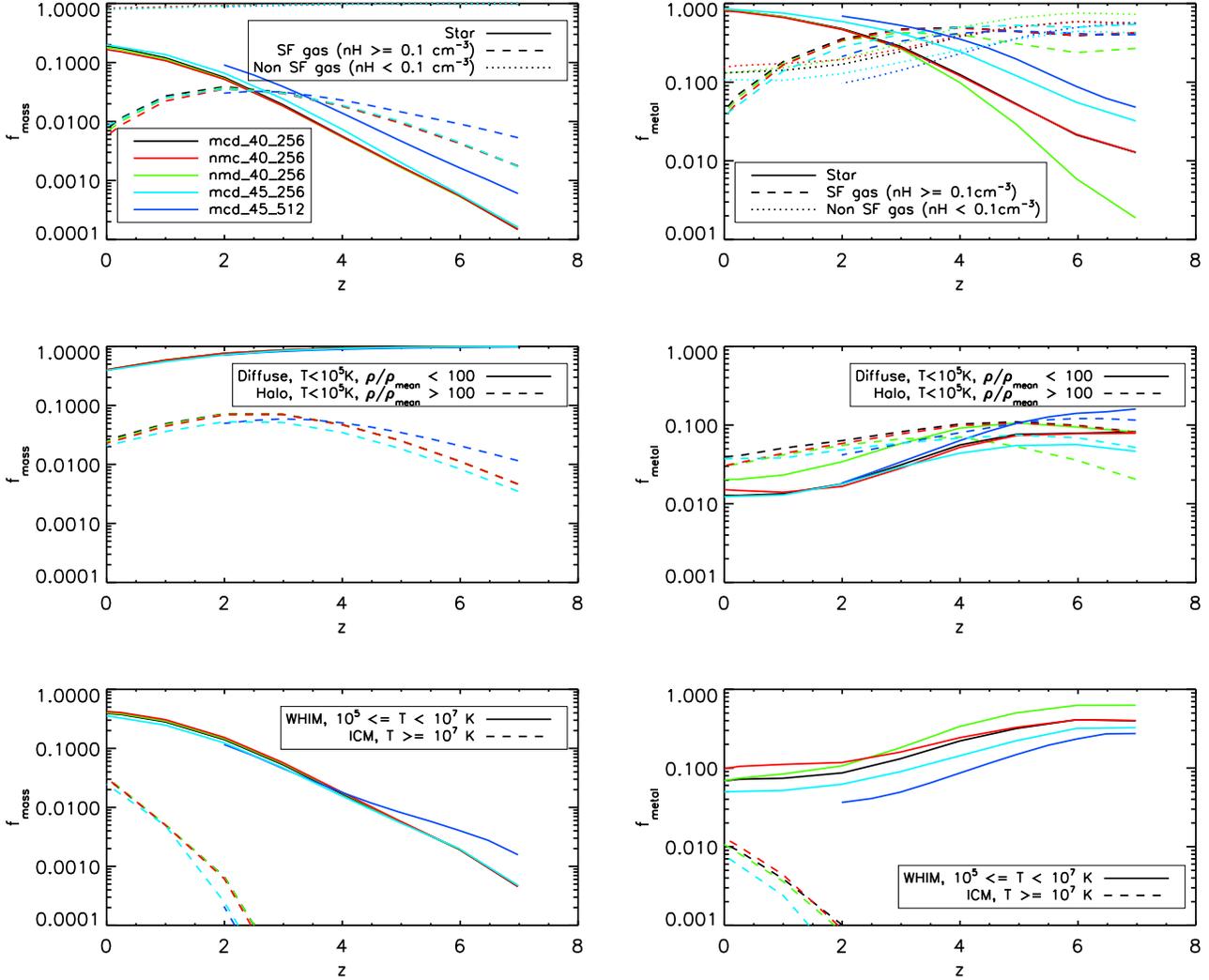}
\caption{The evolution of baryon mass and metal fraction in various
  gas phases 
  and stars. The gas is divided as star forming gas,
  diffuse IGM, halo gas and WHIM according to the definition described
  in the text. {\it  Black}: the standard run ``mcd\_40\_256''. {\it
    Red}: the run with  
  metal cooling turned off.  {\it Green}: the run with no metal
  diffusion. {\it Cyan}: medium resolution run in the convergence test
  ``mcd\_45\_256''. {\it Blue}: high resolution run ``mcd\_45\_512''.
  The meaning of each line type in each panel is defined by the
  legends in that panel. } 
\label{fractions}
\end{figure*}

Figure~\ref{fractions} shows how the gas and metal fractions
evolve. The left panels focus on 
the mass fractions in gas and stars.  Almost all baryonic mass is in  
the diffuse IGM at z=7.  By z=0, the mass fractions in 
stars, the diffuse IGM and the shock heated WHIM are about 20\%, 40\%
and 40\%, respectively for the moderate resolution runs.  The stellar mass fraction at z =0 is
consistent with observations \citep[][and references
therein]{Wilkins08}, although close to the high end.  A large amount of
WHIM forms at low redshift when 
gas is shock heated when falling into halos. The mass fraction of WHIM
gas is consistent with SPH simulations using explicit 
galactic superwind models \citep{Oppenheimer06, Choi09}, and Eulerian 
simulations \citep{Cen06b}. The amount of star forming gas (ISM) and
galactic halo gas evolves similarly to the SFH shown in Figure
\ref{figsfr}, reflecting a close relation between the ISM, halo
gas and SF activity. 

The blue and cyan lines in Figure \ref{fractions} show the
  convergence of mass fractions.  At high redshift, the $512^3$
  simulation has a larger mass fraction in stars, the ISM, halo gas
  and the WHIM. Resolving smaller halos at large z increases the
  amount of gas in galaxies (halos and the ISM) and enhances SF.
  Consequently, enhanced stellar feedback increases the amount of
  WHIM.  This effect is more significant at higher redshift and
  decreases with time. The WHIM fraction converges at z $<$ 4. The ISM
  and halo gas fractions reach  similar values as in the low resolution
  run at z =2, although they seem to decline slightly faster towards
  low z.  For the stellar mass, however, there is about 37 \% more in
  the high resolution run at z = 2. Although it is still difficult to
  predict the stellar fraction at z = 0,  we expect that the
  difference will be lower at z = 0 since there is already no excess in the reservoir gas (the ISM) at z =2.  In fact, early star formation seen in the high resolution simulation may consume the gas that would otherwise contribute to SF at low redshift, hence the SFR may decline faster towards low z, which may imply a rapid convergence in stellar mass fraction.

The right panels of Figure~\ref{fractions} show how the
metal fractions present in stars and different gas phases evolve. At z
= 7,  gas (SF and Non-SF) contains the majority of the metals while
stars contain only a few percent, which is possibly because the time
scale for gas consumption is longer than 
  the age of the Universe, so that metals in the ISM do not 
  have enough time to be incorporated into stars. With metal diffusion, initially the
  non-SF gas (the IGM + halo gas) and the SF gas (i.e. the ISM) contain comparable amounts of metals, with the former having slightly more in most runs. With time, the metal fraction in the non-SF gas decreases and the metals budget is dominated
  by the ISM and later by stars. As z=0, 80\% to 90\% of metals reside
  in stars. The IGM and halo has about 10\% of the metals and the ISM
  has only a few percent.  The result that the IGM contains a larger
  fraction of metals at higher redshifts suggests 
  that the enrichment process is very efficient at early epochs, 
  possibly because it is easier for 
wind material to leave the shallow potential wells of early objects.
For different phases of the IGM, metal fractions in both the WHIM and
the diffuse IGM decrease with time. For example, in the standard run,
the WHIM metal fraction decreases from $\sim$ 40\% at z=7  to about
10\% at z=0, the metal fraction in diffuse IGM decreases from $\sim$
10\% to $\sim$ 1\%. The ICM, on the other hand, increases its share of
metals as galaxy clusters form at low redshift.  
 
Metals in the IGM are primarily in the WHIM, with a smaller fraction in
the diffuse IGM. This remains the case for the entire
cosmic history, which differs from \citet{Dave07}, who found
that  most metals reside in the cooler diffuse IGM from $1.5 < z < 6$.
This is probably because they use hydrodynamically decoupled,
 kinetic feedback.  Their
superwinds were generated in cold SF gas and hydrodynamically decoupled
when they left the ISM, and the wind material was likely to be
cool. Our feedback model ejects SN energy into the surroundings and
suppresses the gas cooling and the winds are generated by thermal
pressure, which makes them more likely to be hot and 
in the WHIM phase. \citet{Wiersma09} obtained  a similar result as ours
  for the IGM metals using a local energy injection and 
hydrodynamically coupled wind model from
\citet{DallaVecchia08}. However, metals in their 
simulations mostly reside in the ISM at redshift  z $>2$ while in our
results the IGM generally contains more metals at high redshift. While
more controlled simulations are necessary to explain the difference,
it seems that subgrid feedback models alter metal enrichment
significantly.   

The high resolution simulation increases the metal fraction in stars
while decreasing it in the ISM and Non-SF gas. Between different
phases of the Non-SF gas, halo gas and the diffuse IGM increase their shares 
of metals, while metal fractions in the WHIM decreases. This is
expected since with high resolution SN feedback events are more
frequent but each has smaller impact. As a result, particles that receive feedback
energy and metals generally have shorter cooling
shut-off times in the blastwave model \citep{Stinson06}. Thus they can be
cooled more efficiently by metal cooling to the diffuse IGM, or to the
cool halo gas. No clear convergence is seen at z = 2 when the high resolution
simulation was terminated. The differences in stars, the ISM, halo gas and the WHIM
are 17\%, 25\%, 14\% and 41\%, respectively. However the trend of
metal fraction evolution as a function of redshift remains similar for high and
low resolution runs. The general conclusions about enrichment
efficiency and about metal distribution in each phases also remain the same.

\subsection{Evolution of Metallicity}

\begin{figure*}
\includegraphics[width=175mm]{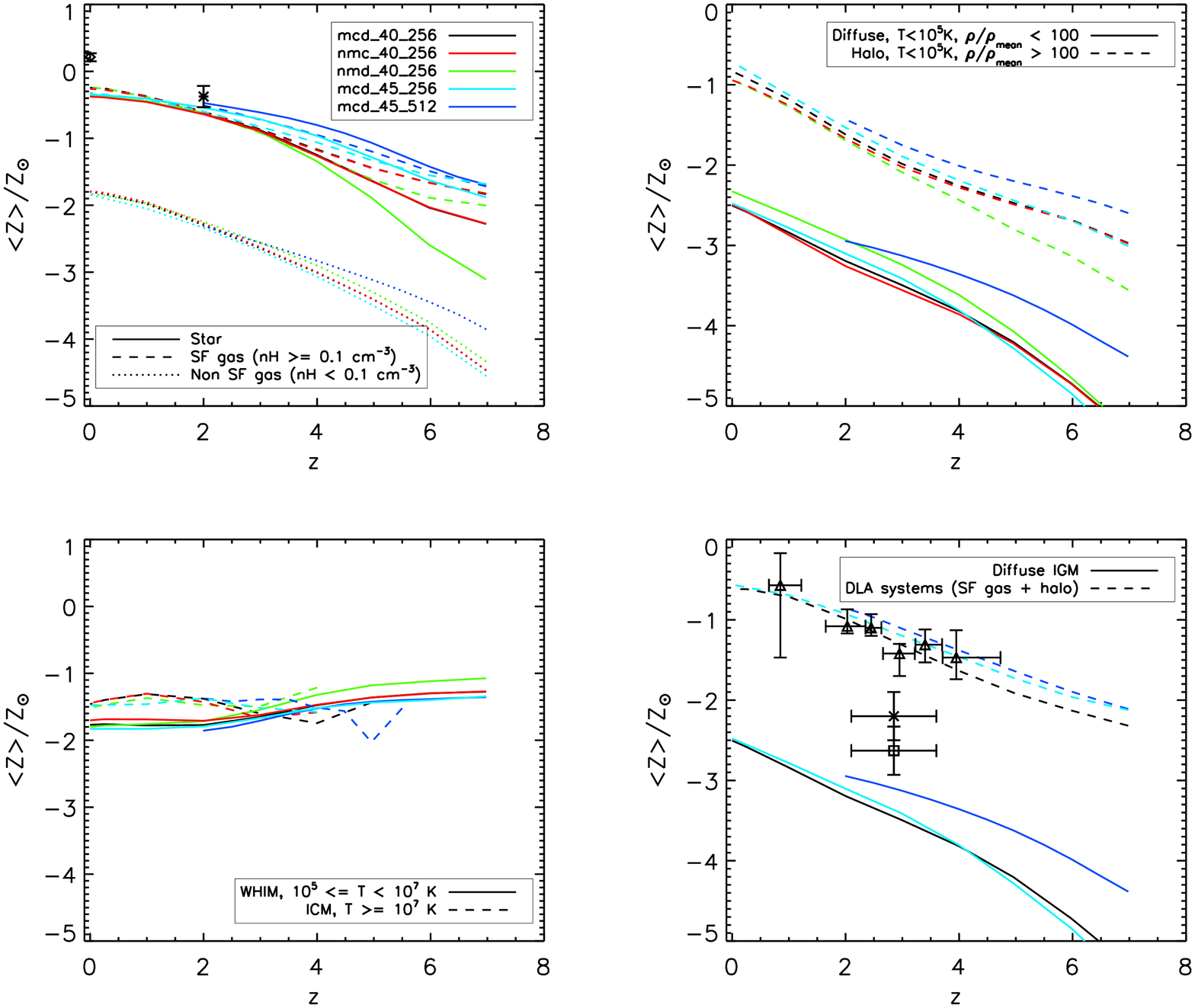}
\caption{The evolution of metallicity in stars and various gas phases
  described in the text.  The same legend as in Figure~\ref{fractions}
  is used. The solar metallicity is defined as $Z_{\sun}$ =
  0.0127. Observational data: {\it Diamond}: stellar metallicity at z
  = 0 \citep{Gallazzi08}.  {\it Asterisk}: stellar metallicity at z = 2
  from \citet{Halliday08}. {\it Triangles}: the metallicity evolution
  of DLA and sub-DLA systems from \citet{Prochaska03}.  {\it Cross}:
  the IGM metallicity traced by O VI from \citet{Aguirre08};  {\it
    Square}: the IGM metallicity traced by C IV from
  \citet{Schaye03}. The observations were scaled to the same solar abundance as the simulations. Z$_{\sun}$ = 0.0127  } 
\label{metallicity}
\end{figure*}

Figure~\ref{metallicity} shows how the mass weighted metallicity in
stars and the gas phases evolves. Stars and SF gas have the highest
metallicities throughout the simulation. The metallicities are about
$10^{-2}$ Z$_{\sun}$  at z = 6-7  and steadily increase to $\sim$ 0.5
solar at z=0. At z =0, the value is in agreement with the observed
stellar metallicities in galaxy groups, Z$_{\star} \sim$ 0.6 Z$_{\sun}$
\citep{Finoguenov03} and simulation results from \citet{Dave07}.
However, it is about 0.5 dex 
lower than the measurement from \citet{Gallazzi08} (diamond symbol in
left upper panel of Figure~\ref{metallicity}), although our total
stellar metal density ($\Omega_{\rm Z}$) is 4-5$\times 10^{-5}$ at z
= 0, consistent with the \citet{Gallazzi08} value.  At z = 2,  the
stellar metallicity is consistent, but slightly lower than the 
observation by \citet{Halliday08}. 

Metallicities of stars and SF gas are similar throughout the
simulation.  The non-SF gas has an overall similar trend of evolution
as the stars and SF gas, but is less enriched on average.  The
metallicity of the diffuse IGM evolves 
significantly but is much less enriched comparing to other phases. At
z=0 it is only $10^{-2.5}$ Z$_{\sun}$.  The metallicities of the WHIM
and the ICM evolve slowly with a slight decreasing trend, despite
their rapid mass increase at low redshift. Both values vary 
between 0.01 $Z_{\sun}$  and 0.1 $Z_{\sun}$.  Since WHIM and the cool
diffuse IGM contains similar mass at z = 0, the fact that WHIM
metallicity is much higher than the diffuse gas implies our IGM
enrichment process happens primarily in the WHIM. The convergence
is better for metallicity than metal fractions, and the metallicities of
most gas phases are nearly converged at z = 2. The largest difference
is seen for diffuse IGM and the halo gas, where the high resolution
run has about 43\% and 26\% higher metallicites, respectively. It
  may be because feedback gas has a shorter cooling shut-off time so
  it is more likely to cool to $< 10^{5}$ K, or because the high
  resolution run can better resolve star formation in smaller halos at
  higher redshift, and the metals ejected from those objects have not
  been heated above $10^{5}$ K. 

We computed the mean metallicities of the halo gas  
and the ISM and compared it with the metallicity evolution of the DLAs
and sub-DLAs from \citet{Prochaska03} (triangle symbols in the bottom
right panel of Figure~\ref{metallicity}).  The DLA and sub-DLA systems are
considered tracing the gas in galactic disks and halos. Our results
compare well with observations in this case. We also compare the
diffuse IGM metallicity with observations using C~IV (\citet{Schaye03},
square symbol) and O~VI (\citet{Aguirre08}, cross symbol) at z $\sim$
2.5. Our results are smaller than both observations even for the
high resolution case. Since the results do not converge here,
  this discrepancy could be due to a lack of
  resolution. Alternatively, it may also suggest additional feedback
  mechanisms which can enrich diffuse IGM are probably necessary
  here. For the intracluster medium,  
observations found higher metallicities than our results at z=0, about 
0.2 -0.5 Z$_{\sun}$ \citep[][and references therein]{Aguirre07}. 
This inconsistency may be because of the AGN feedback absent in the
simulations. Moreover, the size of our
simulation also limits the number of galactic clusters that can form
(below the cosmic average).  Since our simulations were designed for studying
metal enrichment in the IGM,  we will not further address the
intracluster medium.

\subsection{The Effects of Metal Cooling and Metal Diffusion}
\label{diffuse_cool1}

\subsubsection{The Effects of Metal Cooling}

The red curves in the left panels of Figure~\ref{fractions}
show the evolution of mass fraction without metal cooling. At z = 0,
metal cooling increases the stellar and halo mass fraction  
each by $\sim$ 14\%, increases the SF gas fraction by 32\%, and decreases
the WHIM fraction by 7\%. It reflects that metals enhance the cooling 
significantly at WHIM temperatures ($10^{5} K - 10^{7} K$) and thus
more WHIM cools onto galaxies and forms stars. The mass of diffuse IGM is
almost unaffected by metal cooling ($\sim$ 1\% decrease at z =0)
as its metallicity is very low, as shown in Figure \ref{metallicity}.   

The right panels of Figure~\ref{fractions} show that the
metal cooling decreases the metal fraction in the WHIM
and increases 
it in the halo gas and the SF gas, because it enhances the cooling of enriched
WHIM onto galactic halos and disks.  At z=0, the decrease 
in WHIM gas is $\sim$ 24\%, the increase in halo gas is $\sim$
42\%. For stars and SF gas the increments are 14\% and 32\%,
respectively.  For the WHIM and the halo gas, the effects of cooling in
metal fractions (which traces the enriched gas) are larger than in mass
fractions (which traces the total gas).  For star and SF gas, the
effect of metal cooling is similar in both mass and metal fractions.
For the diffuse IGM, metal cooling increases its metal fraction at z
$> 1.5$, but decreases it at lower redshift.  The early increase is
likely due to the cooling of WHIM to become diffuse IGM, while the later
decrease is probably because of the diffuse IGM itself is enriched
enough so that metal cooling can enhance its accretion onto halos.   

\subsubsection{The Effects of Metal Diffusion}

Metal diffusion produces similar effects as metal cooling on mass fraction.
  Metal diffusion contaminates large amounts of otherwise 
pristine gas with metals, enabling it to cool through metal lines
from WHIM to SF gas or halo gas,  and enhances the SFR.  As it 
depends on the metal cooling, the effects of metal diffusion on mass fraction are
smaller in magnitude than having metal cooling 
off (similar to the effects on the SFH). At z = 0, the increase in
halo and stellar mass fractions are 
about 7\% and 5\%, respectively,  and the decrease in WHIM is $\sim$ 3\%.    

The green curves in the right panels of Figure~\ref{fractions}
show the direct impact of metal diffusion on the metal content of
various gas phases.  Metal diffusion increases the
metal content of stars, the ISM and halo gas (i.e., gas in the
galaxies), and decreases the metal content of the WHIM, the ICM and
the diffuse IGM (i.e., the IGM). This is counter intuitive since one
expects diffusion to distribute the metals more evenly. However, as
the enriched gas is ejected out of the galaxies, metal diffusion
mixes its metals with the ambient gas (i.e., the ISM and the galactic
halos) along the outflow trajectory so that by the time the outflow
reaches the IGM (WHIM, diffuse 
gas or the ICM),  its metal content has decreased. In other words, diffusion
prevent highly enriched gas from transporting all its metals to the
WHIM. This is further discussed in Section~\ref{phasediagram}. Metal diffusion
may alleviate the problem found in previous SPH 
simulations where the metals are too inhomogeneous
\citep[e.g.,][]{Aguirre05}. We will make a detailed 
analysis of the effects of metal diffusion in observable ions such as C
III, C IV and O VI in a future paper.  

The impact of metal diffusion changes with redshift.  It is more
significant at high redshift for the WHIM, the halo, the ISM and the
stars. For example, from z = 7 to z = 0, the increase in star metal
fraction varies from 0.4\% to 580\%, in SF gas from 7\% to $\sim$ 60\%, 
in halo gas from 30\% to 300\%. The decrease in the WHIM metal
fraction varies from 0.6\% to 60\%.  Since diffusion arises from 
velocity shear, the diffusion effects are high near winds.  The
concurrent decrease in the WHIM metals and the increase in stellar,
SF or halo metals imply that the winds take metals from galaxies
directly to the WHIM. The fact that the diffusion impact is higher at
high redshift suggests that the winds between these phases are more
effective at early times. 
 
\begin{figure*}
\includegraphics[width=85mm]{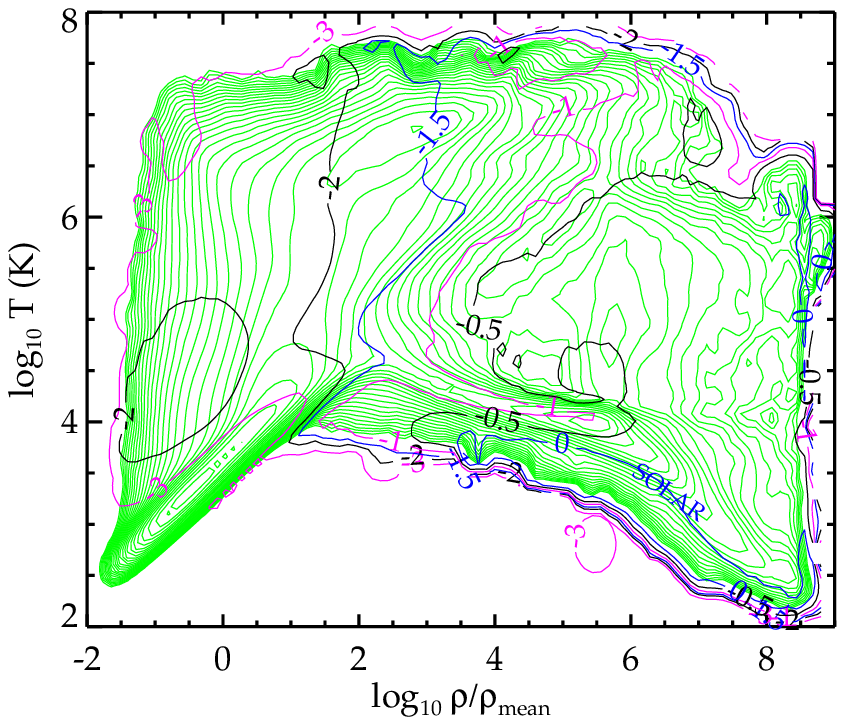}
\includegraphics[width=85mm]{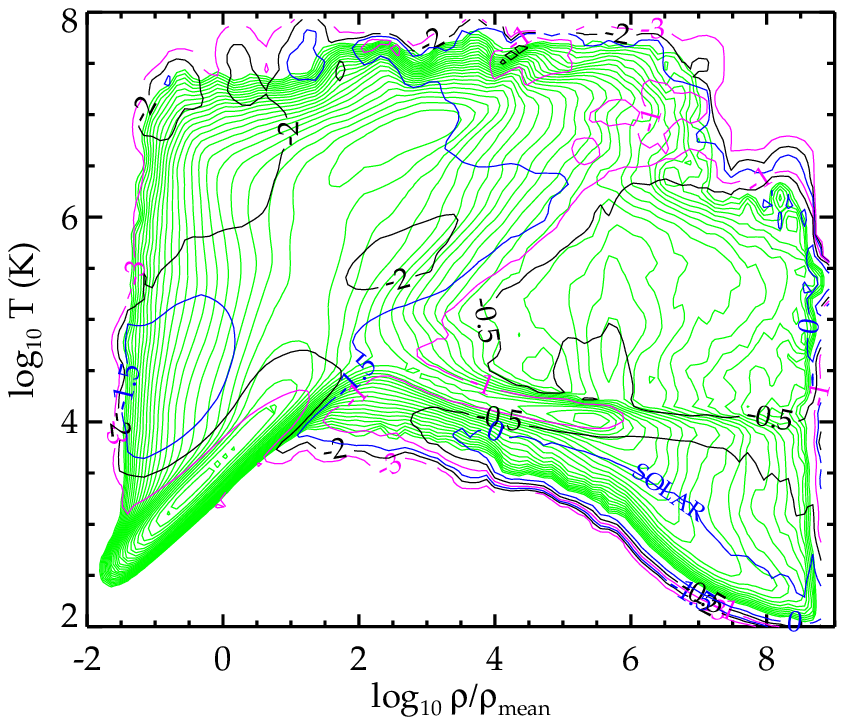}
\caption{Distribution of Gas and Metallicity in the $\rho$-T phase
  diagram of the reference simulation at z=0.  The green contours
  indicate the gas mass  
  distribution, and the black,magenta and blue contours with labels indicate the
  metallicities. From low to high, the metallicity labels are [Fe/H] =
  -3.0, -2.0, -1.5, -1.0, -0.5 and 0.0.  Solar metallicity is
  defined as Z$_{\sun}$ = 0.0127.  The left panel is the standard run
  with metal diffusion and the right panel is without diffusion.
}
\label{rhotz}
\end{figure*}

\section{Distribution of Gas and Metals in Density and Temperature at z=0}
\label{phasediagram}

In this section, we examine the detailed distribution of mass and
metals in the density-temperature ($\rho$-T) phase diagram at z=0. In
Figure~\ref{rhotz}, the green contours indicate the mass
distribution of gas across the phase diagram while the labeled darker
contours indicate mean metallicities for that gas.  The left panel shows
the standard run while the right panel has diffusion turned off.
Looking at the standard run, it can be seen that gas with T
$< 10^{4} K$ and $\rho < 10\ \rho_{mean}$ follow a power law equation
of state (EOS). It is heated by photoionization
and cooled by adiabatic expansion and has generally not participated
in star formation or feedback, following the standard expectation for
the diffuse IGM \citep[e.g.][]{Hui97}.  At $\rho > 10\ \rho_{mean} $, the gas
distribution splits into high and low temperature branches. The low
temperature branch extends to star-forming gas in galaxies.  Due to
metal cooling, gas with $\rho > 10^{4} \rho_{mean}$ can reach
temperatures below the atomic hydrogen cooling cut-off at $\sim 10^4$
K.  The WHIM is apparent as the less dense gas with temperatures above
$10^{5}$ K. There is less gas around $\sim 10^{5.5}$ K compared to the
peaks at $10^{4}$K and $10^{6.5}$K due to the peak in metal cooling rates at
$10^{5}$ to $10^{6}$ K. 

\begin{figure}
\includegraphics[width=\columnwidth]{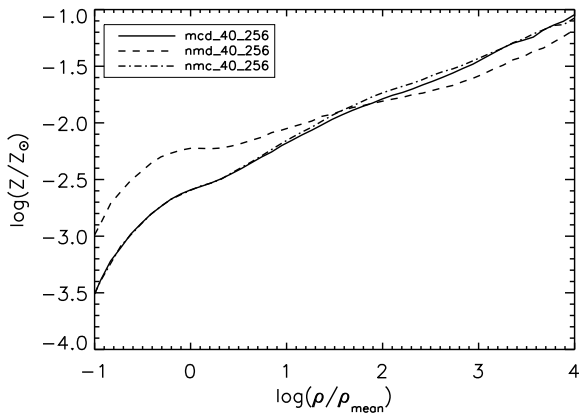}
\caption{The metallicity-density relation at redshift zero for all the
  simulation runs. The effect of metal diffusion is shown in the
  excess at the low over density range and the decrease of metallicity
  in the halo gas.}
\label{rho_zmetal}
\end{figure}

The WHIM arises from both metal enriched winds and pristine virial
shocks.  Wind material can get out to very low densities but with
realistic diffusion operating the typical metallicities are not that
high and the cooling times can be very long.  The most enriched gas is
at the highest densities: star forming gas in galaxies.  The
metallicity of this gas can be super-solar.  The highly enriched
gas (near solar metallicities) is all above $\rho/\rho_{mean} $
$\sim10^{4}$.  The metallicity decreases as the wind propagates and
mixes moving to lower densities.  This steady progression with
density can be seen in the hot gas ($T > 10^5$ K).  The
WHIM is enriched up to $10^{-2}$ Z$\sun$ to $10^{-1}$ Z$\sun$, while the diffuse
IGM following the power-law EOS is barely enriched. The metallicity
contours in Figure \ref{rhotz} show that metal enriched IGM is hotter than
this diffuse phase.       

The metallicity distribution on the phase diagram without metal
diffusion in the right panel of Figure~\ref{rhotz} is similar, but the
metals are less 
evenly distributed.  In particular, more low density material becomes
enriched.  For example,  some gas around T$\sim 10^{4} K$ and
$\rho/\rho_{mean} \sim 0.01$  has metallicities up to 0.1 solar. On the other
hand,  relatively dense, cool halo gas ($\rho/\rho_{mean} >100$, T$<
10^{5} $K) is less enriched,  as shown in the plot that the horizontal
strip of lower metallicities extends to higher densities. As discussed
in Section 
\ref{diffuse_cool1}, this is because without metal diffusion, the
metals are locked into the original wind material and must travel with it.

The overall trend in metallicity versus density can be seen in
Figure~\ref{rho_zmetal}.  The effect of metal diffusion can be clearly
seen.  The reference run produces a positive, near linear relation
between the density and metallicity in logarithmic space for
$\rho/\rho_{mean} > 1$.  The slope is larger for the under-dense
gas. The simulation without metal diffusion increases the under-dense
gas metallicity by a factor of 3 while it decreases it for halo gas
(with overdensities larger than 100).  Overall, it gives a shallower
slope in the range of log$(\rho/\rho_{mean})$ = [0, 3].  The result
from the simulation without metal cooling is also plotted, although
the effect is relatively small.   Observationally, the
metallicity-density relation can be inferred from certain metal
tracers in the QSO spectra. For example, \citet{Schaye03} used pixel
statistics of C IV and found the carbon abundance follows a log-linear
relation with density at the higher redshifts (z = 1.8-4.1).  These
observations may strongly constrain the feedback and wind generation
mechanism, but uncertainties due to the shape of the UV background and
the ionization states of the metal tracers may change the results
substantially, as discussed in \citet{Schaye03} and
\citet{Oppenheimer06}.  We defer detailed comparison of specific
species with observations to a following paper.

\begin{figure*}
\includegraphics[width=175mm]{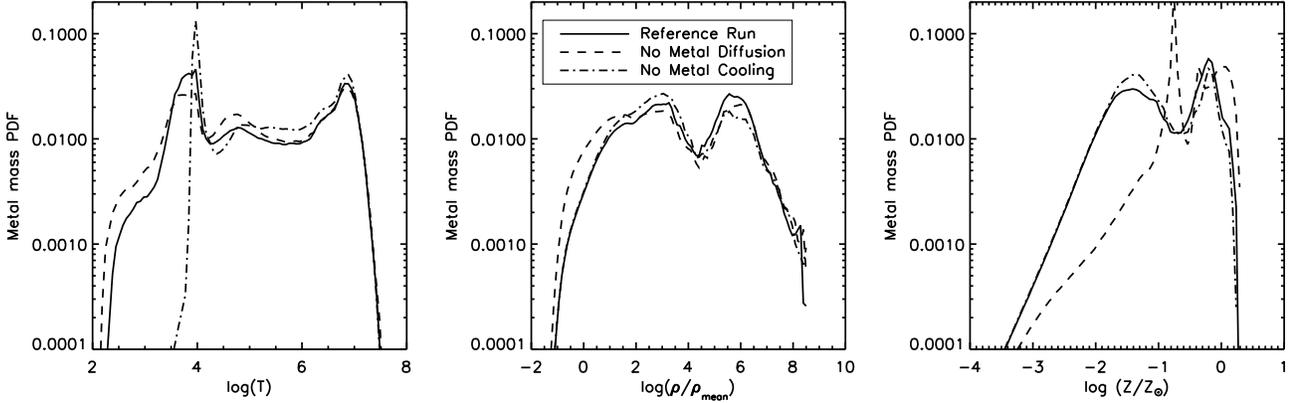}
\caption{The probability density function (PDF) of the metal mass over
  temperature (left panel), density(center panel) and metallicity
  (right panel) at z=0.  {\it Solid lines}: the reference simulation;
  {\it Dashed lines}: the simulation without metal cooling; {\it
    Dot-dashed lines}: the simulation without metal cooling. } 
\label{mzpdf}
\end{figure*} 

Figure~\ref{mzpdf} shows the metal-mass-weighted probability density
functions (PDF) over temperature, density and metallicity. Bimodal
distributions are seen in the PDF of all three variables. 
As seen in the leftmost panel, metal cooling is efficient at WHIM temperatures
and shifts a substantial amount of material from the WHIM to lower 
temperatures and the sharp peak of gas at
$10^{4} $K associated with primordial cooling is removed.   
Without diffusion, super-enriched gas is present which is able
to cool to nearly 100 K.   Without diffusion, enriched gas is also able to 
travel to extremely low densities as seen in the second panel.

The impact of metal diffusion is clear in the metallicity distribution
(the rightmost panel in Figure~\ref{mzpdf}). With metal diffusion,
there is a large increase in material at low metallicites, Z $<$ 0.1
Z$_{\sun}$).  Without diffusion the metals concentrate in a spike
just above 0.1 Z$_{\sun}$ associated with the early distribution of
metals from a star formation event. Turbulent mixing spreads the 
metals from the wind material to the surroundings so that
substantially more gas is contains metals at low levels. An
increase in low metallicity gas was also achieved in \citet{Wiersma09}
using a local smoothing technique, but a physically motivated
diffusion model results in the metals being substantially
redistributed in space as seen in the second panel of the figure.
 
\section{Characterizing the Wind Mechanism}
\label{wind}

The primary difference between the simulations presented here and
earlier enrichment studies is the mechanism for wind generation.
Earlier simulations explicitly added velocity to gas in regions where
stars form.  Here feedback only added thermal energy and prevented gas
cooling according to recipe developed using simulations of isolated
Milky Way like disks and dwarf galaxies.     
We have shown in previous sections that our adiabatic feedback can
cause mass loss from
galaxies to enrich the IGM. However, since winds are produced
dynamically as a result of  
feedback instead of via a wind recipe, it is helpful to
characterize the wind generation mechanisms. In this section, we
examine mass loss as a function of halo mass for our baseline
simulations. The purpose is to understand how our current adiabatic
feedback generates winds that enrich the IGM.   Our results do
allow scope for  more vigorous feedback.  Additional feedback
such as explicit superwinds and subgrid AGN models may be
included in our future enrichment studies. The results presented here
serve as a starting point for further investigations testing more
complicated feedback scenarios.

\begin{figure}
\includegraphics[width=85mm]{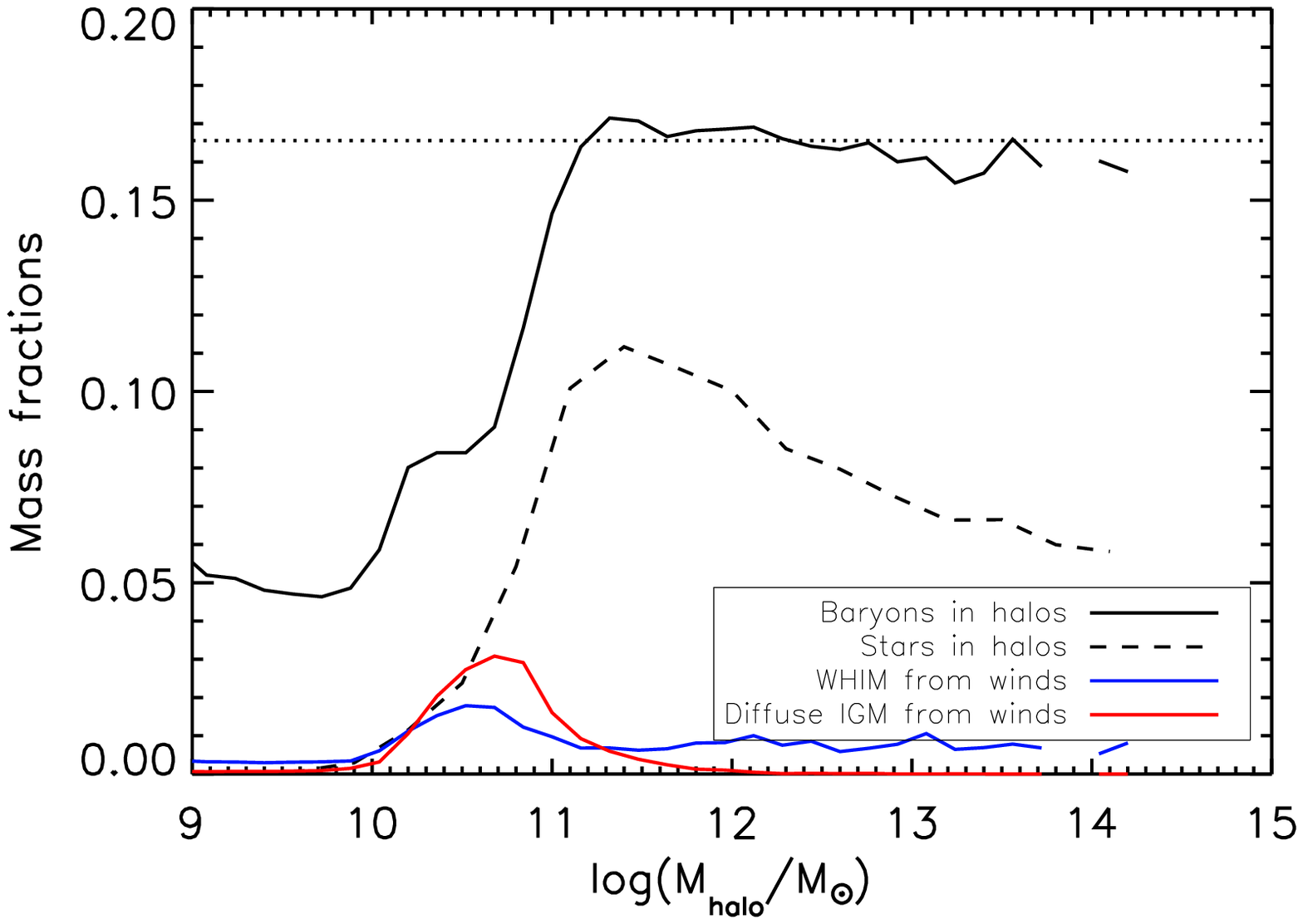}
\includegraphics[width=85mm]{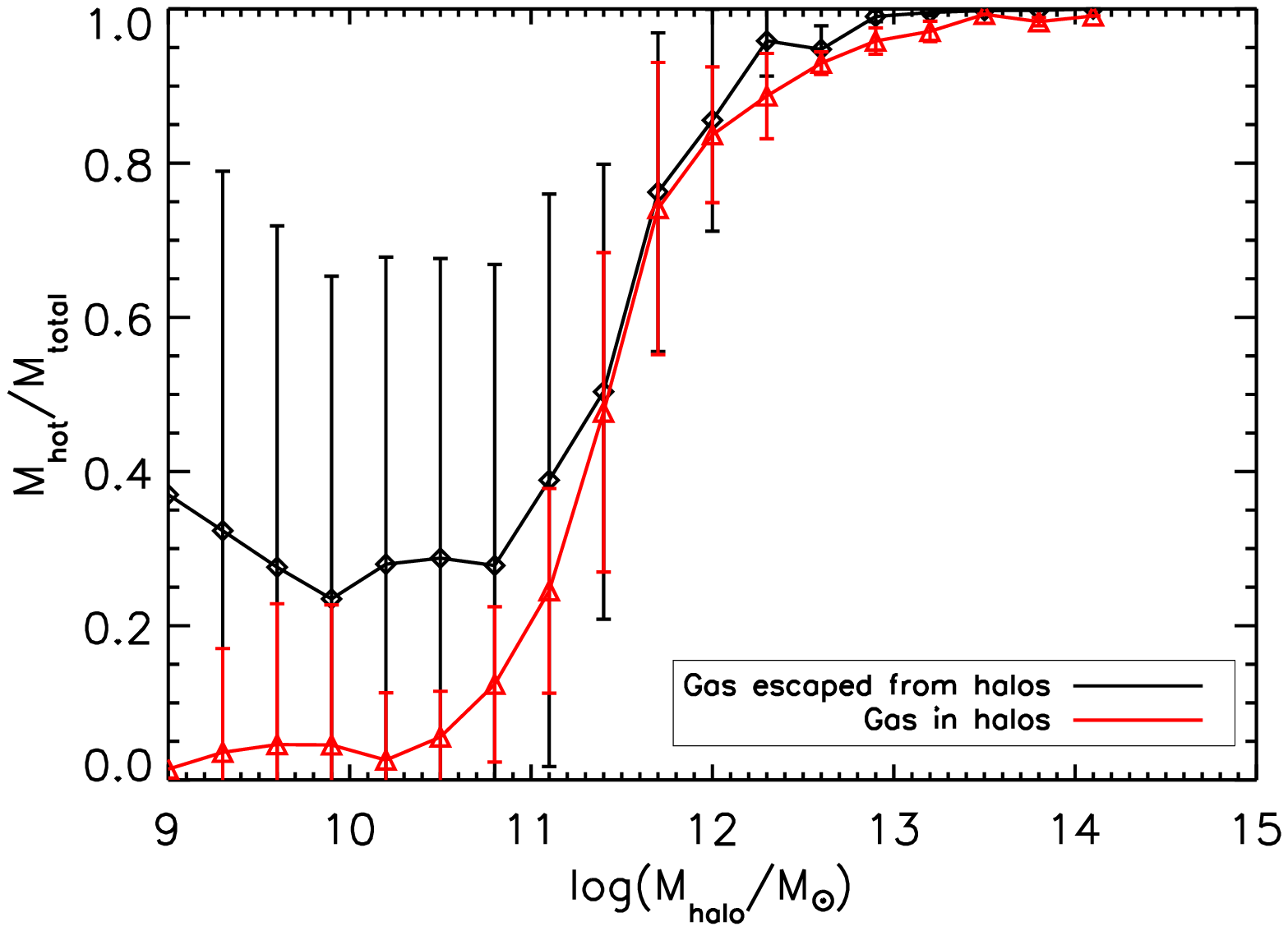}

\caption{{\bf Upper panel:} Distribution of baryon and stellar mass
  fractions of halos and wind fraction escaped from halos,
  as a function of total halo mass in the reference run (z=0). {\it
    Black solid line:} baryon fraction (include gas and stars) in
  halos. {\it Dotted line}: the cosmic mean value. {\it Dashed line}:
  stellar mass fraction. To indicate how halos of different mass lose
  gas and the role of stellar feedback and winds, we identify the lost
  gas by paring each gas particle with its dark matter neighbour in
  the initial condition, and finding the ones who have their DM
  partners within halos but themselves in the IGM. When plotting this
  gas as a function of halo mass, we mean the mass of halos that its
  pairing dark matter particle is belong to. The gas found in
  this method is further split to the WHIM phase ($10^{5}$ K $< $ T $<
  10^{7}$K) and the cooler IGM phase (T $< 10^{5}$K). Among this gas, winds were
  identified by tracing the density history of the gas using
  $\rho(z)_{max}/\rho_{mean} > 100$ and $\rho(z=0) < 0.5\
  \rho(z)_{max}$. {\it Blue solid line}: mass fraction of the
    halo baryons that escaped as wind and are currently in the
    WHIM. {\it Red solid line}:  mass fraction of the halo
    baryons that escaped as wind and are currently in the diffuse
    IGM.  {\bf Lower panel}: For the gas that is currently
    within or outside of a certain halo, the mass fraction of this gas
    that is in the warm-hot phase, as a function of the halo mass. 
  {\it Black curve with error bars}: mass fraction of the gas outside of
    halos (identified by the partner DM particles) that is in
  the warm-hot phase (i.e. WHIM). {\it Red curve with error bars}:
  mass fraction of halo gas that is in the warm-hot phase. } 
\label{baryondis}
\end{figure}

We identified overdense halos using a friends-of-friends (FOF) group finder
with linking length $\epsilon$ = $\frac{1}{5}$ (inter-particle
separation $\sim 30$ kpc) to find material
inside regions of overdensity $\frac{\delta\rho}{\rho} \ga$125.       
We compared the baryon fraction ($M_{bary}/M_{tot}$) of material inside
the halos (black solid line) as a function of total halo mass at z = 0 with       
the cosmic mean ($\Omega_{b}/\Omega_{m}$, dotted line) in the upper panel of 
Figure~\ref{baryondis}.   The baryon content is low for low mass halos ($<
10^{10} M_{\sun}$).  It increases quickly with mass in the
intermediate range, and stays near the cosmic mean beyond $
\sim 10^{11} M_{\sun}$.  Halos of masses less than $\sim 10^{11} M
_{\odot}$ contain less than the cosmic mean. Using a lower resolution
($2 \times 128^{3}$ particles) run, we ruled out that the decrease below
$\sim 10^{11} M_{\odot}$ is a resolution effect. This decrease
indicates that gas was either prevented from accreting onto those halos, or 
ejected/stripped from them. Possible mechanisms that cause the mass
loss include: 
1. Background UV radiation heating the gas and preventing it from
accreting onto dwarf galaxies. 2. Tidal stripping during
merger events. 3. Winds generated from the stellar feedback mechanism.

We investigated where the ``lost'' baryons are at the current epoch by
pairing each gas particle with its dark matter (DM) partner at the
same location in the initial conditions.  We 
identified the gas particles  that remain in the IGM (i.e. not belong
to any groups) even though their 
partner dark matter particle has accreted to a halo at z=0. When
plotting any properties of this gas as a function of halo mass, we
mean the mass of halos to which the DM partner of the gas is belong.
Note that this method only considers the difference between the
  initial condition and the snapshot at z = 0, so it does not capture
  the information in between such as re-accretion of gas after
  ejection. Hence the result here is used as a qualitative indication
  of the fate of the ``lost'' gas. A more precise description would require
  track gas particles.  According to its
  temperature, this ``lost''  gas was categorized as WHIM ($10^{5}$K $<$ T $ < 10^{7}$K ) or cooler, diffuse gas (T $< 10^{5}$K). The bottom panel of
  Figure~\ref{baryondis} shows the mass fraction of this gas that is
  in the form of WHIM (black line with
  error bars).  For comparison, the gas inside halos was also
  categorized in the same way as warm-hot halo gas and cool halo gas,
  and the fraction of the halo gas that is in the warm-hot form
  is shown in the same panel (red line with error bars).  Despite of
  the large error bars,  the curves show that gas associated with smaller halos 
  tends to be cooler than that associated with large halos. This is
  expected because larger halos have larger gravitational potentials
  and higher virial temperatures.   
 
We investigated the role of stellar feedback and winds in the efficient
mass loss by tracing the
density history of the ``lost'' gas up to z = 5 to identify winds. If
its highest overdensity, $\rho(z)_{max}/\rho_{mean}$,  was 
larger than 100, and its current density is less than half of the
maximum ($\rho(z=0) < 0.5\ \rho(z)_{max}$), then the gas is considered
ejected as wind material. Since the gas density was measured only
  when snapshots were generated every $\Delta$z = 0.25, it is possible that
  some wind gas was omitted. However, 
we verified this density method in the non-diffusion
simulation,  where enriched winds can be identified by the
 metallicities of the particles.  Our density method successfully
detects the same material identified by metals.  The blue
and red lines in the upper panel of Figure~\ref{baryondis} indicates 
fractions of the halo baryons that were expelled as winds, and are
  currently in the WHIM and cool IGM phase, respectively. We also
  plotted the stellar mass fractions in the same figure (dashed line). 

Our results show that baryon loss due to winds is most efficient at
intermediate mass ranges, $10^{10}$ to $10^{11} M_{\sun}$. This is
also the range where the stellar fraction increases rapidly, which
reflects the correlation between star formation and wind generation. Below
$10^{10} M_{\odot}$, baryon fractions of the halos are very low and
there are almost no stars and no winds. We traced the accretion
history of this gas and found it has never accreted onto any
objects. Though higher resolution is necessary 
to investigate these objects, it is likely that heating due to the
background UV radiation prevents gas accreting onto such halos, since UV
heating can bring gas temperature to $10^{4}K - 10^{5}K$, comparable
to the virial 
temperature of dwarf galaxies with mass $\la 10^{10} M_{sun}$. Wind
fractions start to decrease when M $ > 10^{10.5}$M$_{\sun}$. 
This may be caused by 1. decrease of wind escaping efficiency
with increasing halo gravitational potentials; or 2. decrease of star
formation efficiency as halo mass increases, as indicated in the
decline of stellar fraction in Figure~\ref{baryondis}. The latter
is probably due to the increase of cooling time, as the virial
temperature of the halo increases and the cooling rate decreases beyond $10^{6}$ K
(ref. Figure~\ref{figcooling}). For halos with mass $\gg 10^{12}$M$_{\sun}$, virialized gas
can have longer cooling time than the dynamical timescale and hence fuel supplies 
for SF activities can be largely reduced \citep{Rees77}.  It is worth
noting that the stellar fraction from our simulation is 
larger than observations \citep[e.g.,][]{Guo09} so there is an
``overcooling'' problem, in which gas in simulations tends to cool too
rapidly and form much more stars comparing with observations. It was
suggested that feedback from Active Galactic Nuclei (AGN) may
alleviate this problem \citep[e.g.][]{McCarthy09} especially for
galactic groups. However, a discussion of AGN feedback is beyond the
scope of this paper, and we use the stellar fraction curve only to
indicate SF activities in different mass halos.

The colored lines in the upper panel of Figure
\ref{baryondis} show that the wind material can be in WHIM or cooler
IGM phases. Galaxies of all masses larger than 
$10^{10}$M$_{\sun}$ generate winds in the form of WHIM, while only
galaxies that in range of $10^{10}-10^{12} M_{\sun}$ have cool
wind gas.  We tracked the temperature history for this cool wind gas and
found that about 60 \% of it had $T > 10^{5} K$ when it was generated and has 
subsequently cooled. 

\begin{figure}
\includegraphics[width=\columnwidth]{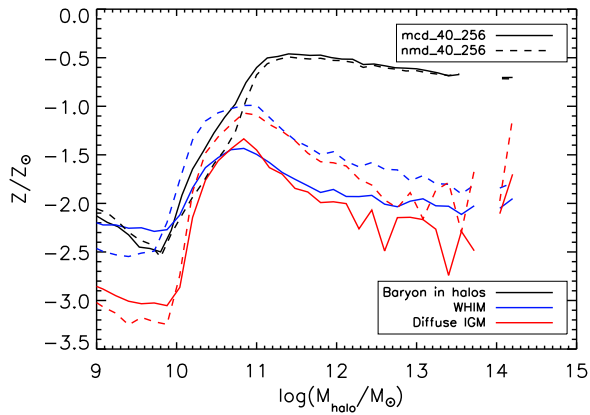}
\caption{The distribution of metallicities of baryons within halos and
  different phases of gas that cannot accrete onto or escaped from halos   
  (described in the caption of Figure~\ref{baryondis}) as a function of
  total group mass at z = 0.  The solar metallicity used in this plot
  is Z$_{\sun}$ = 0.0127. }
\label{zmetaldis}
\end{figure}

Figure~\ref{zmetaldis} shows the metallicity distribution for the
baryons within halos, WHIM and diffuse gas as functions of halo
mass. Note that here the WHIM and the diffuse gas shown are those
  that have their paired DM particles inside halos.  It is the gas that
  cannot accrete or has escaped from halos. This is a subset of the
  total WHIM and diffuse gas described in Section \ref{mass_metal_evo}. The 
metallicity for all gas is low for halos with mass $\la
10^{10}$M$_{\sun}$.  For the diffuse IGM and the WHIM, it is likely because
this gas has never accreted onto galaxies and experienced star 
formation. For the gas in halos, the metallicity is still low due
to low star formation. In the intermediate mass range,
the metallicity of the gas in halos increases significantly,
because the baryon content of halos increases and the galaxies
undergo multiple star formation events and accumulate metals.
Consequently, winds are enhanced, increasing the metallicities in the
diffuse and WHIM gas outside those halos. The metallicity of baryons
in halos saturates above  $\sim 10^{11}$M$_{\sun}$,
reflecting the decline of SF activity with increasing halo mass.  The
metallicities of the escaped material (in both the WHIM and cooler IGM
phase) decreases, possibly due to the inefficiency of winds leaving
more massive objects and the decline of star formation.  Note that the
amount of diffuse gas is 
very small in massive objects beyond $10^{12}$M$_{\sun}$ so the
metallicities have large uncertainties. 

The dashed lines in Figure~\ref{zmetaldis} show the metallicities of
different gas phases for the non-diffusion run. The effects of metal
diffusion are seen in two aspects. First, metal diffusion contaminates
the gas that is prevented from accreting due to UV or stripped by
tidal forces, as indicated in the increases of the metallicity of the
gas associated with low mass halos when metal diffusion is on.
Second, metal diffusion allows winds to lose metals along the wind
trajectory, therefore significantly decreasing the metallicity in
winds, while increasing it in halo gas.  This is reflected in the
metallicity decrease of the gas outside halos (both the WHIM
and the diffuse gas) at mass larger than $10^{10}$M$_{\sun}$,  and the
increase in the metallicity of the baryons within halos.  Without
diffusion almost all metals in the unbound gas are locked in wind
material (98\% of all metals in the WHIM and 99\% of
all metals in the diffuse IGM). In the full model, including metal diffusion, the numbers
decrease to 58\% and 80\%, respectively.  

\section{Summary and Conclusions}
\label{summary}

We investigated the enrichment of the intergalactic medium with SPH
cosmological simulations using an adiabatic stellar feedback model.
The simulations incorporated a self-consistent metal
cooling model with an ultraviolet (UV) ionizing background along with
metal diffusion that models the turbulent mixing in 
the IGM and the ISM.  It was found that the UV background significantly alters the
metal cooling rates at all temperatures from 100 K to $10^{9}$
K.  Above $10^{4}$ K it decreases the cooling rate and shifts the cooling
peak to higher temperature, while below $10^{4}$ K the UV increases the
metal cooling rates due to the increase of free electrons.

The simulations produced an SFH broadly consistent with observations to redshift z $\sim$ 0.5,  and a steady
cosmic total neutral hydrogen fraction ($\Omega_{\rm H I}$) that compare relatively well with observations, although possible
discrepancies in the observed $\Omega_{\rm H I}$ at z $\sim$ 2-3 allow
for more vigorous mass-loss.  This demonstrates that adiabatic
feedback can moderate SF while maintaining a regular supply of H I. The evolution of the mean flux decrement
in the Ly-$\alpha$ forest in our simulations is consistent with observations to z $\sim$ 3-4, if the magnitude of the UV background is lowered with respect to the \cite{HM05} rates in \textsc{cloudy}. 

As the universe evolves, there is a rapid increase in the amount of
warm-hot intergalactic medium (WHIM) and a decrease in the cooler 
diffuse IGM.  At z=0, about 40\% of the mass is in WHIM, consistent previous simulations with different methods \citep{Cen06b, Oppenheimer06}. The metal content of the Universe evolves from the largest fraction being in the
IGM to the majority residing in star forming gas, to ultimately being locked in stars at the present day. These trends
reflect more effective wind escape at high redshift.  IGM metals primarily reside in the WHIM, 
unlike \citet{Oppenheimer06} and \citet{Dave07}, whose metals largely
reside in cool gas, as the wind gas starts cool with kinetic feedback
and less likely to be heated up when the wind material is decoupled
from the ISM.  Our result is however in agreement with
\citet{Wiersma09} who also used kinetic feedback but with
non-decoupled wind models.  
%Compared to the observations, our simulations produce a similar 
%estimated metal budget at z=0.  At z $\sim$ 2-3, our results are consistent 
%with observations for stellar and IGM (diffuse plus WHIM) metal fractions, 
%but have significantly more metals in the star forming gas.   
The mean metallicities of stars, star forming gas, galactic halo gas and the
cold diffuse IGM all increase with time, but those of the WHIM and
the ICM remain mostly constant with a slight decreasing trend. The
metallicity of the WHIM stays between 0.01 to 0.1 $Z_{\sun}$. The metallicity of the ICM is similar
to the WHIM, which is smaller than the observed value, 0.2-0.5 $Z_{\sun}$ \citep{Aguirre07},
possibly due to absent AGN feedback.   The metallicity evolution of
the gas in galaxies compares well with DLA and sub-DLA observations
from \citet{Prochaska03} and the metallicity of the diffuse IGM at z $\sim$ 2.5 is
less than the observations from \citet{Schaye03} and
\citet{Aguirre08}, suggesting higher resolution, or additional
mechanisms for enriching the diffuse IGM are probably necessary. 
 
We characterized our galactic mass-loss and wind generation.  For the
current adiabatic feedback model, winds are most efficient for
galaxies in the intermediate mass range of $10^{10}$M$_{\sun}$ to
about $10^{11}$M$_{\sun}$.  Below $10^{10}$M$_{\sun}$ gas is likely to be prevented from
accreting due to UV heating and remains as
low-metallicity gas.  Above about $10^{11}$ M$_{\sun}$ fraction of
wind gas decreases, possibly because the wind escape efficiency
decreases with increasing halo potentials, or because the decline of
star formation activities especially for massive halos. Most winds were hot when generated, 
but the ones expelled from intermediate mass range 
galaxies, having temperatures $\sim 10^{5} - 10^{6}$K, could
cool through metal lines and become diffuse IGM rather than WHIM. 

We investigated the effect of metal cooling and diffusion on the SFH,
the evolution of $\Omega_{\rm H I}$ and the evolution of mass and metals
in different phases.  For metal diffusion, we further studied its
effect in the density-temperature phase diagram at z =0 and the distribution of baryons as function
of the galaxy mass.   Metals significantly enhance the cooling of the
WHIM, allowing the gas to cool and join galactic disks.  Metals also enable cooling
below $10^{4}$K.  Metal cooling decreases the mass and metal fractions
of the WHIM while increasing the metals in stars, halos and SF gas and
increasing the SFR by 20\% and $\Omega_{\rm H I}$ by 17\% at z =0.
With realistic diffusion included, metals mix between winds and surrounding gas
before they leave the galaxies, decreasing the metal content
in the WHIM and diffuse IGM but 
increasing it in the galactic halo and star forming gas. It prevents
enriched, hot winds from creating highly-enriched low density
regions, and makes the density-metallicity relation smoother so it 
follows a nearly log-linear relation in the density range
$log(\rho/\rho_{mean}) = 0-4$.          

We performed two additional simulations with one of which having
  8 times more particles,  to investigate the convergence of our results
  and the resolution dependence of our model.  Full convergence is not
  seen in most of our results, however the results from the high
  resolution run in SFH, $\Omega_{\rm H I}$ evolution, Ly$\alpha$
  decrement evolution are still broadly consistent with observational
  data, and the differences caused by resolution decrease with
  redshift. Also, the basic results of metal fraction evolution trend
  and enrichment efficiency in different gas phases remain unchanged.
  The impact of resolution can be seen in three aspects. Firstly, with
  high resolution the amount of neutral gas (H I) and SFR increase,
  hence the stellar mass fraction and the WHIM mass fraction related
  to wind activities.  The difference is largest at high redshift.
  Secondly, the metal fraction locked up in stars increases with
  resolution while decreasing that in the ISM and IGM in general,
  suggesting a decline 
  in wind efficiency. Thirdly, with high resolution the diffuse IGM and
  halos have a higher metal fraction at the cost of the WHIM
  metals.

In future work, we will make a more detailed comparison with metal
absorption line observations.  We will also 
examine the detailed behaviour of various wind models and the effect
on the immediate environment of galaxies.

\section*{Acknowledgements} 
We would like to thank Robert Wiersma for many useful discussions
during the writing of this paper. Andrew Hopkins and Steve Wilkins
kindly provided their star formation history data      
converted for the Kroupa IMF used in this paper. 
The simulations were performed using the Shared Hierarchical Academic
Research Computing Network 
(\textsc{sharcnet}) facilities.  This work is supported by the
National Science and Engineering Research Council (\textsc{nserc}) of
Canada and a Canadian Institute for Theoretical Astrophysics
(\textsc{cita}) National fellowship.

\end{document}